\documentclass[aps,prl,twocolumn,floatfix,english,showpacs,10pt,superscriptaddress]{revtex4-1} \pdfoutput=1
\usepackage{amsmath,bm}
\usepackage{mathtools}
\usepackage{commath}
\usepackage{amssymb}
\usepackage{graphicx}
\usepackage{babel}
\usepackage{cases}
\usepackage[colorlinks=true, citecolor=blue, anchorcolor=green]{hyperref}
\hypersetup{linkcolor=red}
\usepackage{natbib}
\usepackage{floatrow}
\usepackage{dblfloatfix}
\usepackage{xcolor}
\usepackage{braket}

\makeatletter

\newcommand{\Rmnum}[1]{\expandafter\@slowromancap\romannumeral #1@}
\newcommand{\kd}{\text{1D}}
\newcommand{\ex}{\text{ex}}

\newcommand{\eff}{\text{eff}}

\makeatother

\begin{document}

\title{Subradiant emission from regular atomic arrays: universal scaling of decay rates from the generalized Bloch theorem}
\author{Yu-Xiang Zhang}
\email{iyxz@nbi.ku.dk}
\affiliation{Niels Bohr Institute, University of Copenhagen, Blegdamsvej 17, 2100 Copenhagen, Denmark}
\author{Klaus M{\o}lmer}
\email{moelmer@phys.au.dk}
\affiliation{Center for Complex Quantum Systems, Department of Physics and Astronomy, Aarhus University, 8000 Aarhus C, Denmark}
\date{\today}

\begin{abstract}
The Hermitian part of the dipole-dipole interaction in infinite periodic arrays of two-level atoms yields an energy band of singly excited states. In this Letter, we show that a dispersion relation, $\omega_k-\omega_{k_\ex} \propto (k-k_{\ex})^s$, near the band edge of the infinite system leads to the existence of subradiant states of finite one-dimensional arrays of $N$ atoms  with decay rates scaling as $N^{-(s+1)}$. This explains the recently discovered $N^{-3}$ scaling and it leads to the prediction of power law scaling with higher power for special values of the lattice period. For the quantum optical implementation of the Su-Schrieffer-Heeger (SSH) topological model in a dimerized emitter array, the band-gap-closing inherent to topological transitions changes the value of $s$ in the dispersion relation and alters the decay rates of the subradiant states by many orders of magnitude. 
\end{abstract}

\maketitle

Subradiance is the phenomenon that radiative emission by an atomic ensemble is  collectively prohibited~\cite{Weiss2018} in contrast to the factor $N^2$ enhancement of the radiation rate by $N$ emitters in the Dicke superradiance~\cite{Dicke1954}. 
The application of the subradiant suppression of radiative decay in quantum memories~\cite{Facchinetti:2016aa,Manzoni:2018aa},
excitation transfer~\cite{Moreno-Cardoner:2019aa,Needham2019aa,Ballantine:2020aa} and topological photonics~\cite{Perczel2017,Bettles:2017aa} has spurred strong interests and a number of results have been obtained that are not yet well understood in a single comprehensive theory. Recently, one-dimensional (1D) emitter arrays with sub-wavelength separations, see Fig.1(a), were found to have subradiant states with decay rates scaling as
$N^{-3}$~\cite{Haakh2016,Tsoi:2008aa,Asenjo-Garcia2017,Albrecht2018,Zhang:2019aa,Yu:2020aa,Dinc:2020aa,Brehm:2020aa}, but examples of rates scaling with $N^{-\alpha}$ with $\alpha>3$ were also soon identified~\cite{Kornovan:2019aa}.

The close relationship between subradiance and the band flatness of collectively shared atomic excitations has been realized to be a crucial component of the collective dipole-dipole interaction~\cite{Poddubny:2020aa}, see also ~\cite{Asenjo-Garcia:2019aa,Jenkins2017,Mirhosseini:2019aa,Guimond:2019aa,
Shahmoon:2017aa,Rui:2020aa,Bekenstein:2020aa,Zhang:2018aa,Ke:2019aa,Schilder:2020aa,Bettles:2019aa,Cremer:2020aa}. 
In this Letter we show that a better understanding of precisely this relationship can explain and predict several characteristics of subradiance.    

\begin{figure}[t]
\centering
\includegraphics[width=0.95\textwidth]{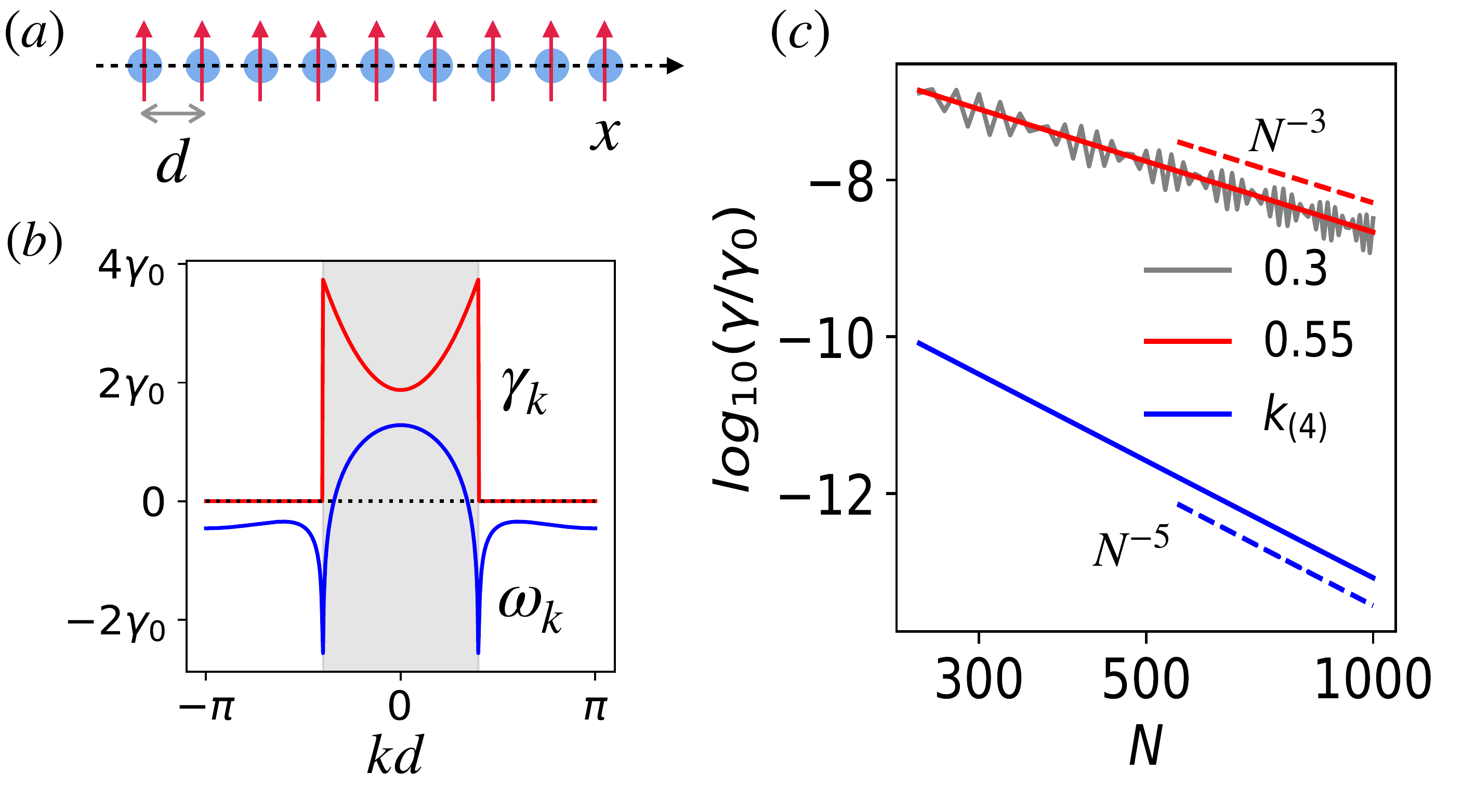} 
\caption{(a) Illustration of a regular array of emitters with dipole moments aligned perpendicular to the spatial array. (b) Energy shifts $\omega_k$ (lower blue curve) and decay rates $\gamma_k$ (upper red curve) 
for the emitter array with $k_0d/\pi=0.4$, where $k_0c$ is the atomic resonance frequency. Wavenumbers outside the shaded interval $\Gamma=[-k_0, k_0]$ correspond to frequencies exceeding the atomic resonance frequency. (c)  Decay rates of the most subradiant states of finite arrays with $N$ emitters in units of the single emitter spontaneous emission rate $\gamma_0$, for $k_0 d/\pi=0.3$ (grey curve), $0.55$ (red curve) and 
$0.4828$ (lower blue curve). The dashed lines show $N^{-3}$ and $N^{-5}$ dependencies.
}
\label{fig1}
\end{figure}

\paragraph*{Dipole-dipole interaction.} 
In regimes where the Born-Markov approximation works well, 
one can trace out the quantized light fields and obtain 
the field-mediated dipole-dipole couplings between the emitters described by an effective Hamiltonian
~\cite{Dung2002}:
\begin{equation}\label{dipole-dipole-3D}
H_{\eff}=-\mu_0\omega_0^2\sum_{m,n=1}^{N}
\mathbf{d}_m^{*}\cdot\mathbf{G}(x_m-x_n,\omega_0)\cdot\mathbf{d}_n
\sigma_m^{\dagger} \sigma_n,
\end{equation}
where $\omega_0$ is the transition frequency
between the emitter ground state $\ket{g}$ and the excited state $\ket{e}$,  $\sigma_m=\ket{g_m}\bra{e_m}$,
$\mathbf{d}_m$ and $x_m$ are the transition
dipole moment and spatial coordinate of the \emph{m}th atom, $\mu_0$ is the vacuum permeability and $\mathbf{G}$ is the dyadic Green's tensor. Our main example is atom arrays along a single dimension in 3D free space, where atoms are equally separated by $d$ and transition dipoles polarized transversally to the lattice direction that depicted in Fig.~\ref{fig1}(a).

Restricting our analysis to the case of a single excitation, shared among the atoms, $H_{\eff}$ is formally equivalent to a non-Hermitian tunneling Hamiltonian among discrete sites $m$, representing the localized excitation, $|m\rangle = \sigma_{m}^{\dagger}\ket{G}$, where $\ket{G}=\ket{g_1 g_2\cdots g_N}$. For an \emph{infinite} array with $-\infty < n,m < \infty$, the dipole-dipole interaction Hamiltonian, $H^{\infty}_{\eff}$  has singly excited right eigenstates in the form of Bloch states, $\ket{k}=\sum_{m=-\infty}^{\infty} e^{ik x_m}\ket{m}$,
with $k \in [-\pi/d, \pi/d]$ and complex eigenvalues $\omega_k-i\gamma_k/2$.
In Fig.~\ref{fig1}(b), the energy shift $\omega_k$ and the decay rate $\gamma_k$ are shown for these states with $k_0=0.4\pi/d$ ($k_0=\omega_0/c$,
$c$ is the speed of light).
Notably, $\gamma_k$ vanishes outside $\Gamma=[-k_0, k_0]$, because the corresponding optical frequencies are not resonant with the atoms~\cite{Asenjo-Garcia2017}. 

In the following we shall make use of the fact that  $H_{\eff} = P_N H^{\infty}_{\eff} P_N$, where $P_N$ projects on the space with no excitations outside the sites $1,2\cdots N$. This implies that the singly excited eigenstates of $H_{\eff}$ can be expanded on the Bloch states,
restricted to the $N$ lattice sites and normalized. We shall refer to these states by the complex argument $z=e^{ikd}$,
\begin{equation}\label{finite-Bloch}
    \ket{z=e^{ikd}}=\frac{1}{\sqrt{N}}\sum_{m=1}^{N} e^{ikx_m}\ket{m}.
\end{equation}
and thus write
\begin{equation}\label{model-Im}
H_{\eff}=N\int_{-\pi/d}^{\pi/d} \frac{\mathrm{d}k}{2\pi/d}(\omega_k - \frac{i}{2} \gamma_k)\ket{e^{ikd}}\bra{e^{ikd}}.
\end{equation}
Note that the states $\ket{e^{ikd}}$ are not orthogonal, and 
hence not the eigenstates of $H_{\eff}$. Therefore, in finite arrays states with $k \notin\Gamma$ are candidate subradiant states with tiny but finite decay rates.

\paragraph*{Generalized Bloch theorem.} To identify the singly excited eigenstates of the finite atomic arrays, the
generalized Bloch theorem~\cite{Alase:2016aa,Cobanera:2017aa,Alase:2017aa}
is essential. The theorem is established for
Hamiltonians in the general form of
\begin{equation}\label{hr}
H_R=h_0\mathbb{I}+\sum_{r=1}^{R}\sum_{m=1}^{N-r} h_r \ket{m}\bra{m+r} + h^{*}_r \ket{m+r}\bra{m},
\end{equation}
where $h_{r}$ are coupling (tunneling) strengths across sites separated by up to a maximum range of $R$. 
$H_R$ is periodic in $m$ except for the leftmost sites $\partial_l=\{1,2,\cdots R\}$ and the, similarly defined, rightmost sites $\partial_r$. 
We denote the projection onto the ``boundary'' $\partial=\partial_l\cup\partial_r$ by $P_\partial$, while the projector on the ``bulk'' sites is denoted by $P_B$ with $P_\partial+P_B=P_N$.

To find eigenstates fulfilling $H_R\ket{\psi}=E\ket{\psi}$, we apply the generalized Bloch theorem noting that the solution space of the \emph{bulk equation} $P_B(H_R-E)\ket{\psi}=0$ is spanned by the states $\ket{z=e^{ikd}}$, where $z$ are the roots  of the equation $\tilde{\omega}_R(z)=E$ with
\begin{equation}\label{Hz-main}
    \tilde{\omega}_R(z)= h_0+\sum_{r=1}^{R} (h_r z^r + h_r^{*} z^{-r}).
\end{equation}
As the array is finite, states $\ket{z}$ with complex $k$ (or equivalently, $\abs{z}\neq 1$) are also physically permitted.  This implies that all the complex roots $z_j$ of the $2R$-degree polynomial equation \eqref{Hz-main}, should be identified. The eigenstate of $H_R$ can then be written as the superposition $\ket{\psi}=\sum_{j=1}^{2R} c_j\ket{z_j}$ that fulfills the boundary conditions, i.e., $P_\partial (H_R-E)\ket{\psi}=0$. 

We note that
Eq.~\eqref{Hz-main} yields the dispersion relation of $H_R$, $\omega_R(k)=\tilde{\omega}_R(e^{ikd})$, and we now suppose that $\omega_R(k)$ has an extremum point $k_{\ex}$ of degree $s$,
i.e., $\omega_R(k)\approx \omega_R(k_{\ex})+a_s(k-k_{\ex})^s$ for $k\approx k_{\ex}$, with $s$ an even integer and $a_s$ the Taylor expansion coefficient. Then
$\tilde{\omega}_R(z)$ can be expanded around $z_\ex=e^{ik_\ex d}$ as
\begin{equation}\label{expansion-ex}
\tilde{\omega}_R(z)= \tilde{\omega}_R(z_{\ex})+a_s\frac{1}{(i d z_{\ex})^s}(z-z_{\ex})^s+\cdots.
\end{equation}
We now focus on 
eigenstates of the finite system with eigenvalues $E\approx\omega_{R}(k_{\ex})$. Since the system has $N$ singly excited eigenstates, it is reasonable to assume that neighbouring states have wavenumbers separated by $O(N^{-1})\pi/d$, and hence a series of eigenvalues may exist with $E=\omega_R(k_\ex)+(a_s/d^s)\delta^s$ where $\delta\sim N^{-1}$. 
Equation~\eqref{expansion-ex} thus yields $s$ roots of $\tilde{\omega}_R(z)=E$ close to $z_\ex$:
\begin{equation}\label{s-solutions}
z_j \approx z_\ex (1+i\delta e^{i2\pi(j/s)}), \quad j=1,2\cdots s,
\end{equation}
while the remaining $2R-s$ roots are not in the vicinity of $z_\ex$.

\paragraph*{A simpler Hamiltonian.} 
We now introduce a Hamiltonian, $\mathbf{H}_{s/2}$, which has its extremum energy at the same $k_{\ex}$ as $H_R$ and a dispersion relation of the same degree $s$, $\tilde{\omega}_{s/2}(z)=\tilde{\omega}_{s/2}(z_\ex)+a_s(-d^2z_{\ex}z)^{-s/2}(z-z_\ex)^s$. $\mathbf{H}_{s/2}$ is chosen such that the roots of $\tilde{\omega}_{s/2}(z)=E$ are given exactly by Eq.~\eqref{s-solutions}. We shall show that the eigenstates of $\mathbf{H}_{s/2}$ approximate the singly excited subradiant eigenstates of $H_{\eff}$ well and permit evaluation of their decay rates by the perturbation theory. 

By introducing $\epsilon_j$ and $\eta_j$ so that $z_j/z_{\ex}=(1+\epsilon_j)^{-1}=1+\eta_j$, we find that the boundary condition implies~\cite{sp}
\begin{equation}\label{conditions}
 \sum_{j=1}^{s} c_j\epsilon_j^r=0,\quad
 \sum_{j=1}^{s} c_j z_j^{N+1}\eta_j^r=0,
\end{equation}
for all powers $r=0,1, 2,\cdots, {s/2-1}$. Eqs.~\eqref{conditions} and the smallness of  $\epsilon_j,\eta_j\sim N^{-1}$ are sufficient to provide effective solutions of the problem without explicitly determining $\{ c_j\}$ and $\{ \epsilon_j, \eta_j\}$.

\begin{figure}[b]
\centering
\includegraphics[width=0.95\textwidth]{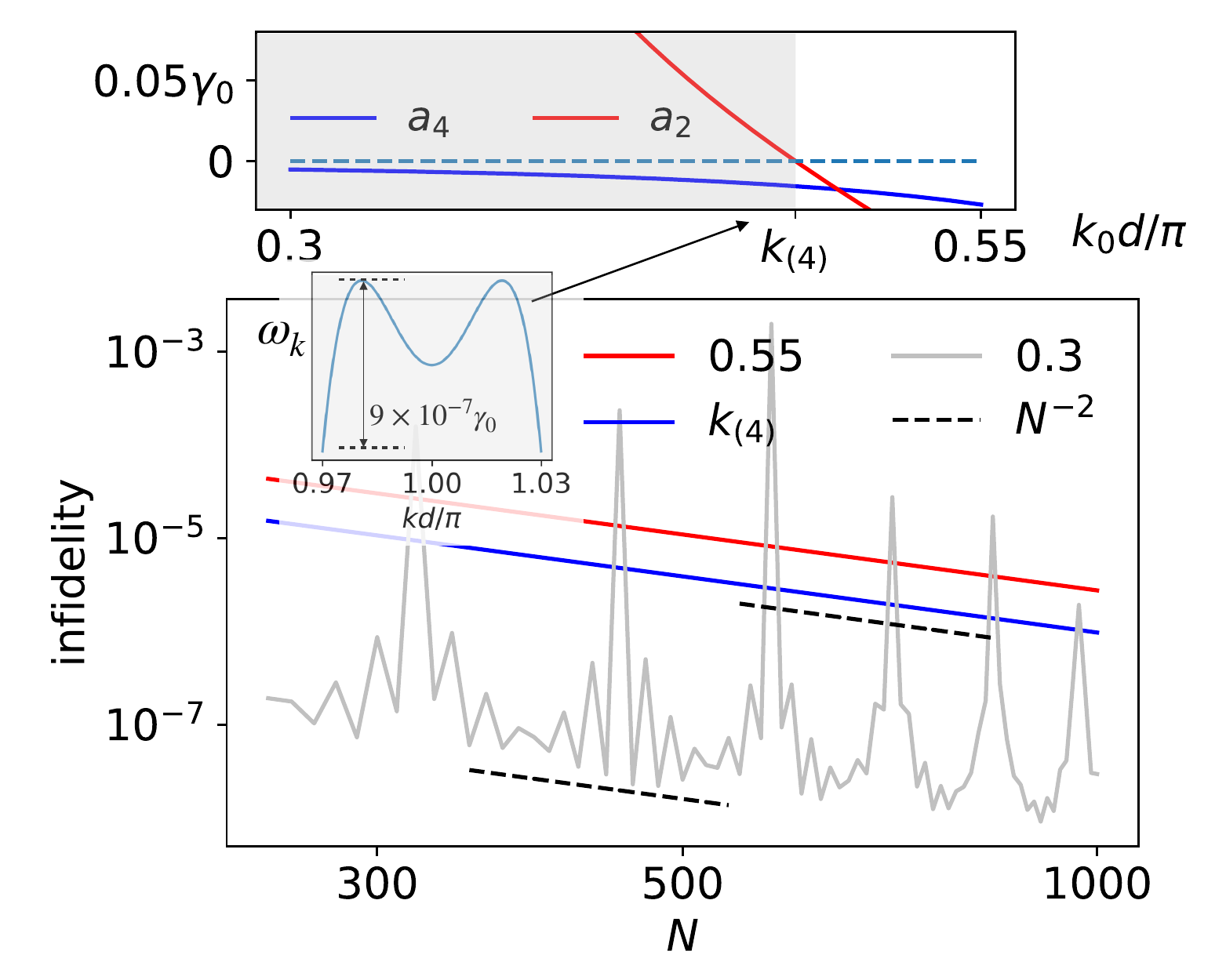} 
\caption{Upper panel: coefficients of the 2nd and 4th order terms of the Taylor series ($a_{2,4}$)
of the dispersion relation around $k=\pi/d$, as a function of $k_0 d/\pi$. 
Lower panel: infidelities (log scale) between the most subradiant right eigenstates of $H_{\eff}$ for $k_0 d/\pi=0.3, 0.55$ and  $k_0=k_{(4)}$, and the eigenstates of $\mathbf{H}_1$ and $\mathbf{H}_2$, respectively. 
The dashed lines indicate the $N^{-2}$ power law behavior.  Insert: the dependence of
$\omega_k$ on $k$ near $k_\ex=\pi/d$, for $k_0=0.4826\pi/d < k_{(4)}$.}
\label{fig2}
\end{figure}

\paragraph*{Perturbative calculation of the subradiant decay rates.} 
While $H_{\eff}$ represented by ${\bf G}(x_m-x_n,\omega_0)$ in Eq.~\eqref{dipole-dipole-3D} does not have a bounded tunneling range, we shall demonstrate that for values of $k$ near $k_{\ex}\notin\Gamma$, 
$H_{\eff}-\mathbf{H}_{s/2}$ can be treated as a perturbation to $\mathbf{H}_{s/2}$. 
The non-Hermitian $H_{\eff}$ can be separated into a coherent part and a dissipative part, $H_{\eff}=H^{\text{Re}}_{\eff}-iH^{\text{Im}}_{\eff}$, {\it cf.}, Eq.~\eqref{model-Im}.
The decay rates of the subradiant eigenstates of $H_{\eff}$ can therefore be approximated by $\gamma=2\braket{\psi|H^{\text{Im}}_{\eff}|\psi}$, evaluated in the eigenstates of $\mathbf{H}_{s/2}$.

Following Eq.~\eqref{model-Im}, we must evaluate $\braket{e^{ikd}|\psi}$ for $k\in \Gamma=[-k_0, k_0]$:
\begin{equation}\label{ka-phi}
\braket{e^{ikd}| \psi}  =\frac{1}{N}\sum_{j=1}^{s}c_j
\frac{z_j e^{-ik d}- (z_j e^{-ik d})^{N+1}}{1-z_je^{-ik d}}.
\end{equation}
Separating the terms in the enumerator and expanding $z_j$ in terms of $\epsilon_j$ and $\eta_j$, we obtain two contributions:
\begin{subequations}\label{derivation}
\begin{equation}\label{key-left}
\begin{aligned}
& \sum\nolimits_{j=1}^{s} c_j\frac{z_j e^{-ik d}}{1-z_{j}e^{-ik d}}
 = \sum\nolimits_{j=1}^{s}c_j \frac{1}{z_{j}^{-1}e^{ik d}-1} \\
= & \frac{1}{z_{\ex}^{-1}e^{ik d}-1}  
\sum_{n=0}^{\infty} \frac{\sum_{j=1}^{s} c_j \epsilon_j^n}{(z_{\ex}e^{-ik d}-1)^{n}},
\end{aligned}
\end{equation}
\begin{equation}\label{key-right}
\begin{aligned}
&\sum\nolimits_{j=1}^{s} c_j \frac{(z_j e^{-ik d})^{N+1}}{1-z_{j}e^{-ik d}} \\
  =& \frac{e^{-i(N+1)k d}}{1-z_{\ex}e^{-ik d}}\sum_{n=0}^{\infty}
  \frac{\sum_{j=1}^{s} c_j z_j^{N+1}\eta_j^n}{(e^{ik d}-z_{\ex})^n},
\end{aligned}
\end{equation}
\end{subequations}
which vanish for $n=0,1,\cdots,{s/2-1}$ due to Eq.~\eqref{conditions}. 

Keeping only the non-vanishing term of the lowest order, $n=s/2$, we  obtain 
\begin{equation}
\begin{aligned}
    \braket{\psi |H^{\text{Im}}_{\eff}|\psi }\leq &
    \frac{1}{N} \bigg(|\sum\nolimits_{j}c_j\epsilon_j^{s/2}|^2
    +|\sum\nolimits_{j} c_j z_{j}^{N+1}\eta_{j}^{s/2}|^2 \bigg) \\
    & \times \int_{-k_0}^{k_0}
    \frac{\mathrm{d}k}{2\pi/d}\frac{\gamma_k}{ |z_{\ex}-e^{ikd}|^{s+2}}.
\end{aligned}
\end{equation}

As $k_{\ex}\notin \Gamma$, the denominator in the integral does not approach $0$, and the integral contributes an $N$-independent finite factor. Using
$\epsilon_j\sim\eta_j\sim N^{-1}$, we thus get the scaling of the decay rate with $N$
\begin{equation}
\gamma=2\braket{\psi |H^{\text{Im}}_{\eff}|\psi }
\sim N^{-s-1}.
\end{equation}
This yields the advertised $N^{-\alpha}$ power law with $\alpha=s+1$. Note that 
$\braket{\psi |H^{\text{Im}}_{\eff}|\psi }$ is a factor $N^{-1}$ smaller than the differences between the real eigenvalues of $\mathbf{H}_{s/2}$ in the vicinity of $\omega_R(k_\ex)$. Thus the perturbation treatment is consistent in the limit of large $N$.

To complete the demonstration, we must also ensure that $\Delta H=H^{\text{Re}}_{\eff}-\mathbf{H}_{s/2}$ can be consistently treated as a perturbation. To this end, we represent $\Delta H$ in the form of Eq.~\eqref{model-Im}, with the dispersion relation $\delta\omega_k=\omega_k-\omega_{s/2}(k)$ and exploit the fact that $\delta\omega_k\sim N^{-s-1}$ for $k\approx k_{\ex}$.
See more details in the Supplemental Material~\cite{sp}.

As a further check of the consistency of our perturbative treatment, we verify that the numerical right eigenstates of $H_{\eff}$, differ by only a small amount from the eigenstates of the simpler Hamiltonian
\begin{equation}\label{infidelity}
\ket{\psi'}\propto \ket{\psi}+O(N^{-1})\ket{\psi^{\perp}}
\end{equation}
yielding an infidelity of,  $1-\abs{\braket{\psi|\psi'}}^2\sim N^{-2}$. 

The $N^{-2}$ scaling of the infidelity is, indeed, confirmed for the subradiant states of our system with decay rates scaling as $N^{-3}$ for $k_0 d/\pi= 0.3$ and $0.55$ (grey and red curves in Fig.~\ref{fig2}), and for the subradiant state with a decay rate scaling as $N^{-5}$ and $k_0 d/\pi= k_{(4)}\approx 0.4828$ (blue curve). We observe that the grey infidelity curve for $k_0=0.3\pi/d$ follows the overall $N^{-2}$ behavior with dramatic oscillations, which are due to an interference effect~\cite{Poddubny:2020aa} between Bloch waves that are degenerate with the extremum of $\omega_k$. This interference is also the cause of the  oscillatory structures in the value of the decay rate as function of $N$ in Fig.~\ref{fig1}(d). The upper panel of Fig.~\ref{fig2} shows the 2nd and 4th order coefficients ($a_{2,4}$) of the Taylor series of $\omega_k$ at $k_\ex = \pi/d$, and  
we see that $a_2>0$ and $a_4<0$ when $k_0 < k_{(4)}$ and hence
band degeneracy is expected, as illustrated in the insert of
Fig.~\ref{fig2}. For $k_0 \geq k_{(4)}$, the extremum is nondegenerate and no oscillations are observed. A similar behavior is displayed in~\cite{sp} for analytically solvable toy model Hamiltonians.

\paragraph*{Qualitative discussion of subradiant decay rates.}
A supplementary, qualitative explanation of why a higher order dispersion relation leads to a higher order $N^{-\alpha}$ decay rate may be inferred from  Fig. 3(b) in Ref.~\cite{Asenjo-Garcia2017}, 
which shows that the radiation from the subradiant states is mostly emitted from the ends of the emitter array. A flat band structure with a larger value of $s$ implies a slower group velocity which extends the excitation lifetime in the system by impeding the propagation of excitation towards the chain ends. 

By the same argument, we expect that subradiant states well inside the energy bands, i.e., in regions of linear dispersion, are characterized by finite group velocities and hence the emission from the ends of the array occur with a rate scaling as $N^{-1}$. 
In conjunction with their numerical discovery of subradiant states with $\sim N^{-3}$ decay rates, Asenjo-Garcia et al.
~\cite{Asenjo-Garcia2017} identified a series of states labelled by an integer $\xi$ and decaying at rates $\sim \xi^2/N^3$.  For $\xi \sim O(N)$, corresponding to wave numbers well inside the energy bands 
($k-k_{ex} \simeq \frac{\xi}{N} \frac{\pi}{d}$~\cite{Zhang:2019aa}), this, indeed, yields decay rates scaling as $N^{-1}$.

Our results imply that varying the power $s$ of the energy band may form practical ways to control the emission of light by emitter arrays. In the remaining part of this Letter, we shall demonstrate such control in emitter arrays that undergo a Su-Schrieffer-Heeger (SSH) type topological transition.

\paragraph*{Dimerized arrays implementing the SSH Hamiltonian.}
We proceed with the study of a dimerized atomic array interacting with the quantized electromagnetic field in 3D free space and in a 1D waveguide. Both systems have topological properties characterized by the SSH model~\cite{Su:1979aa}.
For a recent review on topological Bloch bands, see Ref.~\cite{Cooper:2019aa}. 
Topological transitions are usually accompanied by the closing and opening of gaps in the energy bands. The above analysis suggests that this may radically impact the radiative decay rates of the subradiant states.

\begin{figure}[t]
\centering
\includegraphics[width=0.95\textwidth]{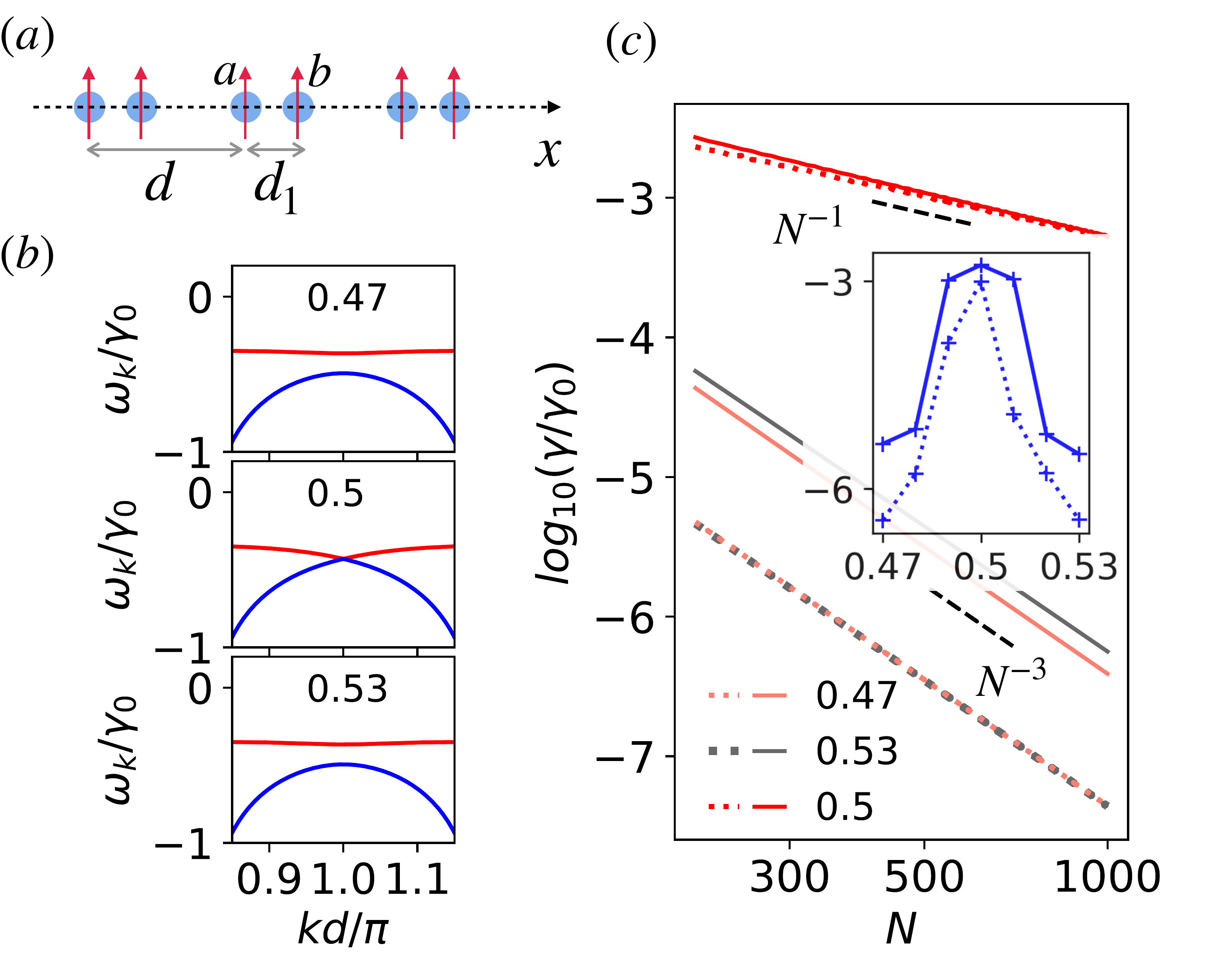} 
\caption{(a) The dimerized array of atomic emitters. 
(b) Dispersion relations for the dimerized array with $k_0 =0.8\pi/d$ and
$d_1/d=0.47, 0.5, 0.53$, respectively. (c) Decay rates of the subradiant states with wavenumbers close to $\pi/d$ as a function of the number of units cells $N$. The dashed lines show the reference $N^{-1}$ and $N^{-3}$ power law dependence for comparison with the numerical results. The insert shows the decay rate as function of $d_1/d$ for $N=500$. The dotted (solid) lines refer to the upper (lower) band.  }
\label{fig3}
\end{figure}

Figure~\ref{fig3}(a) shows the dimerized version of the emitter array, which has the lattice constant $d$ and two atoms (denoted by ``$a, b$'') 
separated by the distance $d_1$ within each unit cell. We denote $d_2=d-d_1$. Two nonequivalent configurations, $d_1<d_2$ and $d_1>d_2$, are found to be topologically trivial and nontrivial
(manifested by boundary states~\cite{Wang:2018aa,Pocock:2018aa}) and the band topology can be 
characterized mathematically by the Zak phase~\cite{Atala:2013aa}. The topological phase transition occurs at $d_1=d_2$, where we recover the regular array in Fig.~\ref{fig1}(a) with the lattice constant $d_1$.
The subradiant states with, e.g., $k=\pm0.5\pi/d_1$ (and $k_0=0.4\pi/d_1$) are well within the regions with linear dispersion, and they have decay rates scaling as $N^{-1}$. The lowest band of the Brillouin Zone of the regular lattice $[-\pi/d_1, \pi/d_1]$ corresponds to two bands of the Brillouin Zone of the dimerized lattice $[-\pi/d, \pi/d]$, where the subradiant states are labelled by $k = \pi/d$ (and where $k_0=0.8\pi/d$). 
To describe the two Bloch bands, Eq.~\eqref{finite-Bloch} 
should be augmented with intra-cell states
\begin{equation}
\ket{e^{ikd},\mathbf{u}^{\pm}}=\frac{1}{\sqrt{N}}\sum_{m=1}^{N} e^{ik x_{m}} \mathbf{u}^{\pm}\cdot\boldsymbol{\sigma}_m^{\dagger}\ket{G},
\end{equation}
where $\boldsymbol{\sigma}^{\dagger}_{m}=(\sigma^{\dagger}_{m,a}, \sigma^{\dagger}_{m,b})$, the 
unit vector $\mathbf{u}^{\pm}=(u_a^{\pm}, u_b^{\pm})$ describes the relative excitation amplitudes inside each unit cell, and ``$+(-)$'' labels the upper(lower) band. As illustrated in the middle panel of Fig.~\ref{fig3}(b),
the two bands of real eigenenergies cross at $k=\pi/d$ with linear dispersion relations.

However, whenever $d_1\neq d_2$, a band gap opens at $k=\pi/d$. This is illustrated in the top and bottom panels of Fig.~\ref{fig3}(b) for
$d_1/d=0.47$ and $0.53$, respectively. When the gap forms, both the upper and lower bands show a quadratic dispersion ($s=2$) around $k=\pi/d$, and we expect the radiative behaviour to change significantly. This, indeed, occurs as evidenced in Fig.~\ref{fig3}(c) where we plot the dependence of the decay rate on $N$ for the subradiant states with wavenumber close to $k_\ex = \pi/d$
for both bands and for the three values of $d_1/d$.
Our numerical calculations clearly show how the $N^{-1}$ dependence of the decay rate for $d_1=d/2$ changes to $N^{-3}$ in case of
$d_1/d=0.47$ and $0.53$. 
A zoom-in on the transition is shown
in the insert of Fig.~\ref{fig3}(c) for the
array emitting into the 3D quantized field with $k_0=0.8\pi/d$ and $N=500$. Notably, the decay rates decrease by three orders of magnitude away from the topological transition. Such critical phenomenon may thus be used to witness aspects of the topological transition.

Analytical results can be obtained for the dimerized arrays coupled to an ideal 1D waveguide. The effective Hamiltonian 
\cite{Chang2012}
\begin{equation}\label{H-waveguide}
H_{\kd}=-i\frac{\gamma_{0}}{2}\sum_{\substack{m,n=1\\ \mu, \nu\in\{a,b\} }}^{N} e^{ik_{0}|x_{m,\mu} - x_{n,\nu}|}
\sigma_{m,\mu}^\dagger\sigma_{n,\nu}.
\end{equation}
has an inverse, $H_{\kd}^{-1}$ that is almost identical to the original SSH model~\cite{Poddubny:2020aa,sp}. Hence
$H_{\kd}$ supports the SSH type topology and the critical points
are found to be $d_1=d_2$ and $d_1=d_2\pm\pi/k_0$.
In~\cite{sp} we focus on the latter values causing 
the band gap opening and closing to occur around $k=0$. At the precise value, $d_1=d_2\pm\pi/k_0$, the subradiant states with wavenumbers close to $k=0$ have decay rates given by~\cite{sp}:
\begin{equation}
\gamma=\frac{\gamma_0}{4N}\cot(k_0 d_1)
\ln(\frac{1+\sin k_0 d_1}{1-\sin k_0 d_1}).
\end{equation}
The $N^{-1}$-scaling of the subradiant decay rates transitions to $N^{-3}$ when $d_1 \neq d_2\pm\pi/k_0$.

\paragraph*{Conclusions.}
We have presented a derivation of a universal  
connection between the decay rates of the most subradiant states of an array of $N$ two level emitters and the Bloch wave dispersion relation near the band edge. This result was demonstrated and explained in detail and it confirms the intrinsic connection between subradiant states and flat energy bands, emphasized in~\cite{Poddubny:2020aa}. We studied the case of radiative emission into the 3D quantized electromagnetic field and a 1D waveguide, but we note that the subradiant phenomena may be further manipulated by coupling to structured radiation reservoirs, such as photonic flat bands~\cite{Leykam:2018aa}. Also, extension of our theory to arrays in two and three dimensions may provide an interesting research area.  

Our study concerned only the linear regime of a single excitation, while we have previously shown that pairs of excitations may survive for even longer times than single excitations in the system. A promising avenue for further research would thus be the exploration of subradiance with many excitations in systems with flat energy bands. Such studies may pose  analogies with phenomena in 
strongly-correlated many-body physics, such 
as, e.g., the fractional Hall effect~\cite{Perczel:2020aa,Parameswaran:2013aa} and the  Lieb lattice \cite{Mukherjee:2015aa,Goldman:2011aa}, see also \cite{Cooper:2019aa,Leykam:2018ab}.

\begin{acknowledgements}
Y.-X. Zhang acknowledges financial support from the Danish National Research Foundation and the European Union’s Horizon 2020 Research and
Innovation Program under Grant Agreement No. 820445 (Quantum Internet Alliance).
K. M{\o}lmer acknowledges support from the Danish National Research Foundation through the Center of Excellence “CCQ” (Grant agreement no.: DNRF156).
\end{acknowledgements}

\bibliography{flatness.bib}

\begin{thebibliography}{48}%
\makeatletter
\providecommand \@ifxundefined [1]{%
 \@ifx{#1\undefined}
}%
\providecommand \@ifnum [1]{%
 \ifnum #1\expandafter \@firstoftwo
 \else \expandafter \@secondoftwo
 \fi
}%
\providecommand \@ifx [1]{%
 \ifx #1\expandafter \@firstoftwo
 \else \expandafter \@secondoftwo
 \fi
}%
\providecommand \natexlab [1]{#1}%
\providecommand \enquote  [1]{``#1''}%
\providecommand \bibnamefont  [1]{#1}%
\providecommand \bibfnamefont [1]{#1}%
\providecommand \citenamefont [1]{#1}%
\providecommand \href@noop [0]{\@secondoftwo}%
\providecommand \href [0]{\begingroup \@sanitize@url \@href}%
\providecommand \@href[1]{\@@startlink{#1}\@@href}%
\providecommand \@@href[1]{\endgroup#1\@@endlink}%
\providecommand \@sanitize@url [0]{\catcode `\\12\catcode `\$12\catcode
  `\&12\catcode `\#12\catcode `\^12\catcode `\_12\catcode `\%12\relax}%
\providecommand \@@startlink[1]{}%
\providecommand \@@endlink[0]{}%
\providecommand \url  [0]{\begingroup\@sanitize@url \@url }%
\providecommand \@url [1]{\endgroup\@href {#1}{\urlprefix }}%
\providecommand \urlprefix  [0]{URL }%
\providecommand \Eprint [0]{\href }%
\providecommand \doibase [0]{http://dx.doi.org/}%
\providecommand \selectlanguage [0]{\@gobble}%
\providecommand \bibinfo  [0]{\@secondoftwo}%
\providecommand \bibfield  [0]{\@secondoftwo}%
\providecommand \translation [1]{[#1]}%
\providecommand \BibitemOpen [0]{}%
\providecommand \bibitemStop [0]{}%
\providecommand \bibitemNoStop [0]{.\EOS\space}%
\providecommand \EOS [0]{\spacefactor3000\relax}%
\providecommand \BibitemShut  [1]{\csname bibitem#1\endcsname}%
\let\auto@bib@innerbib\@empty
\bibitem [{\citenamefont {Weiss}\ \emph {et~al.}(2018)\citenamefont {Weiss},
  \citenamefont {Ara\'{u}jo}, \citenamefont {Kaiser},\ and\ \citenamefont
  {Guerin}}]{Weiss2018}%
  \BibitemOpen
  \bibfield  {author} {\bibinfo {author} {\bibfnamefont {P.}~\bibnamefont
  {Weiss}}, \bibinfo {author} {\bibfnamefont {M.~O.}\ \bibnamefont
  {Ara\'{u}jo}}, \bibinfo {author} {\bibfnamefont {R.}~\bibnamefont {Kaiser}},
  \ and\ \bibinfo {author} {\bibfnamefont {W.}~\bibnamefont {Guerin}},\
  }\href@noop {} {\bibfield  {journal} {\bibinfo  {journal} {New J. Phys.}\
  }\textbf {\bibinfo {volume} {20}},\ \bibinfo {pages} {063024} (\bibinfo
  {year} {2018})}\BibitemShut {NoStop}%
\bibitem [{\citenamefont {Dicke}(1954)}]{Dicke1954}%
  \BibitemOpen
  \bibfield  {author} {\bibinfo {author} {\bibfnamefont {R.~H.}\ \bibnamefont
  {Dicke}},\ }\href {\doibase 10.1103/PhysRev.93.99} {\bibfield  {journal}
  {\bibinfo  {journal} {Phys. Rev.}\ }\textbf {\bibinfo {volume} {93}},\
  \bibinfo {pages} {99} (\bibinfo {year} {1954})}\BibitemShut {NoStop}%
\bibitem [{\citenamefont {Facchinetti}\ \emph {et~al.}(2016)\citenamefont
  {Facchinetti}, \citenamefont {Jenkins},\ and\ \citenamefont
  {Ruostekoski}}]{Facchinetti:2016aa}%
  \BibitemOpen
  \bibfield  {author} {\bibinfo {author} {\bibfnamefont {G.}~\bibnamefont
  {Facchinetti}}, \bibinfo {author} {\bibfnamefont {S.~D.}\ \bibnamefont
  {Jenkins}}, \ and\ \bibinfo {author} {\bibfnamefont {J.}~\bibnamefont
  {Ruostekoski}},\ }\href {\doibase 10.1103/PhysRevLett.117.243601} {\bibfield
  {journal} {\bibinfo  {journal} {Phys. Rev. Lett.}\ }\textbf {\bibinfo
  {volume} {117}},\ \bibinfo {pages} {243601} (\bibinfo {year}
  {2016})}\BibitemShut {NoStop}%
\bibitem [{\citenamefont {Manzoni}\ \emph {et~al.}(2018)\citenamefont
  {Manzoni}, \citenamefont {Moreno-Cardoner}, \citenamefont {Asenjo-Garcia},
  \citenamefont {Porto}, \citenamefont {Gorshkov},\ and\ \citenamefont
  {Chang}}]{Manzoni:2018aa}%
  \BibitemOpen
  \bibfield  {author} {\bibinfo {author} {\bibfnamefont {M.~T.}\ \bibnamefont
  {Manzoni}}, \bibinfo {author} {\bibfnamefont {M.}~\bibnamefont
  {Moreno-Cardoner}}, \bibinfo {author} {\bibfnamefont {A.}~\bibnamefont
  {Asenjo-Garcia}}, \bibinfo {author} {\bibfnamefont {J.~V.}\ \bibnamefont
  {Porto}}, \bibinfo {author} {\bibfnamefont {A.~V.}\ \bibnamefont {Gorshkov}},
  \ and\ \bibinfo {author} {\bibfnamefont {D.~E.}\ \bibnamefont {Chang}},\
  }\href {\doibase 10.1088/1367-2630/aadb74} {\bibfield  {journal} {\bibinfo
  {journal} {New Journal of Physics}\ }\textbf {\bibinfo {volume} {20}},\
  \bibinfo {pages} {083048} (\bibinfo {year} {2018})}\BibitemShut {NoStop}%
\bibitem [{\citenamefont {Moreno-Cardoner}\ \emph {et~al.}(2019)\citenamefont
  {Moreno-Cardoner}, \citenamefont {Plankensteiner}, \citenamefont {Ostermann},
  \citenamefont {Chang},\ and\ \citenamefont
  {Ritsch}}]{Moreno-Cardoner:2019aa}%
  \BibitemOpen
  \bibfield  {author} {\bibinfo {author} {\bibfnamefont {M.}~\bibnamefont
  {Moreno-Cardoner}}, \bibinfo {author} {\bibfnamefont {D.}~\bibnamefont
  {Plankensteiner}}, \bibinfo {author} {\bibfnamefont {L.}~\bibnamefont
  {Ostermann}}, \bibinfo {author} {\bibfnamefont {D.~E.}\ \bibnamefont
  {Chang}}, \ and\ \bibinfo {author} {\bibfnamefont {H.}~\bibnamefont
  {Ritsch}},\ }\href {\doibase 10.1103/PhysRevA.100.023806} {\bibfield
  {journal} {\bibinfo  {journal} {Phys. Rev. A}\ }\textbf {\bibinfo {volume}
  {100}},\ \bibinfo {pages} {023806} (\bibinfo {year} {2019})}\BibitemShut
  {NoStop}%
\bibitem [{\citenamefont {Needham}\ \emph {et~al.}()\citenamefont {Needham},
  \citenamefont {Lesanovsky},\ and\ \citenamefont {Olmos}}]{Needham2019aa}%
  \BibitemOpen
  \bibfield  {author} {\bibinfo {author} {\bibfnamefont {J.~A.}\ \bibnamefont
  {Needham}}, \bibinfo {author} {\bibfnamefont {I.}~\bibnamefont {Lesanovsky}},
  \ and\ \bibinfo {author} {\bibfnamefont {B.}~\bibnamefont {Olmos}},\
  }\href@noop {} {\ }\Eprint {http://arxiv.org/abs/arXiv:1905.00508}
  {arXiv:1905.00508} \BibitemShut {NoStop}%
\bibitem [{\citenamefont {Ballantine}\ and\ \citenamefont
  {Ruostekoski}(2020)}]{Ballantine:2020aa}%
  \BibitemOpen
  \bibfield  {author} {\bibinfo {author} {\bibfnamefont {K.~E.}\ \bibnamefont
  {Ballantine}}\ and\ \bibinfo {author} {\bibfnamefont {J.}~\bibnamefont
  {Ruostekoski}},\ }\href {\doibase 10.1103/physrevresearch.2.023086}
  {\bibfield  {journal} {\bibinfo  {journal} {Physical Review Research}\
  }\textbf {\bibinfo {volume} {2}} (\bibinfo {year} {2020}),\
  10.1103/physrevresearch.2.023086}\BibitemShut {NoStop}%
\bibitem [{\citenamefont {Perczel}\ \emph {et~al.}(2017)\citenamefont
  {Perczel}, \citenamefont {Borregaard}, \citenamefont {Chang}, \citenamefont
  {Pichler}, \citenamefont {Yelin}, \citenamefont {Zoller},\ and\ \citenamefont
  {Lukin}}]{Perczel2017}%
  \BibitemOpen
  \bibfield  {author} {\bibinfo {author} {\bibfnamefont {J.}~\bibnamefont
  {Perczel}}, \bibinfo {author} {\bibfnamefont {J.}~\bibnamefont {Borregaard}},
  \bibinfo {author} {\bibfnamefont {D.~E.}\ \bibnamefont {Chang}}, \bibinfo
  {author} {\bibfnamefont {H.}~\bibnamefont {Pichler}}, \bibinfo {author}
  {\bibfnamefont {S.~F.}\ \bibnamefont {Yelin}}, \bibinfo {author}
  {\bibfnamefont {P.}~\bibnamefont {Zoller}}, \ and\ \bibinfo {author}
  {\bibfnamefont {M.~D.}\ \bibnamefont {Lukin}},\ }\href {\doibase
  10.1103/PhysRevLett.119.023603} {\bibfield  {journal} {\bibinfo  {journal}
  {Phys. Rev. Lett.}\ }\textbf {\bibinfo {volume} {119}},\ \bibinfo {pages}
  {023603} (\bibinfo {year} {2017})}\BibitemShut {NoStop}%
\bibitem [{\citenamefont {Bettles}\ \emph {et~al.}(2017)\citenamefont
  {Bettles}, \citenamefont {Min\'a\ifmmode~\check{r}\else \v{r}\fi{}},
  \citenamefont {Adams}, \citenamefont {Lesanovsky},\ and\ \citenamefont
  {Olmos}}]{Bettles:2017aa}%
  \BibitemOpen
  \bibfield  {author} {\bibinfo {author} {\bibfnamefont {R.~J.}\ \bibnamefont
  {Bettles}}, \bibinfo {author} {\bibfnamefont {J.~c.~v.}\ \bibnamefont
  {Min\'a\ifmmode~\check{r}\else \v{r}\fi{}}}, \bibinfo {author} {\bibfnamefont
  {C.~S.}\ \bibnamefont {Adams}}, \bibinfo {author} {\bibfnamefont
  {I.}~\bibnamefont {Lesanovsky}}, \ and\ \bibinfo {author} {\bibfnamefont
  {B.}~\bibnamefont {Olmos}},\ }\href {\doibase 10.1103/PhysRevA.96.041603}
  {\bibfield  {journal} {\bibinfo  {journal} {Phys. Rev. A}\ }\textbf {\bibinfo
  {volume} {96}},\ \bibinfo {pages} {041603} (\bibinfo {year}
  {2017})}\BibitemShut {NoStop}%
\bibitem [{\citenamefont {Haakh}\ \emph {et~al.}(2016)\citenamefont {Haakh},
  \citenamefont {Faez},\ and\ \citenamefont {Sandoghdar}}]{Haakh2016}%
  \BibitemOpen
  \bibfield  {author} {\bibinfo {author} {\bibfnamefont {H.~R.}\ \bibnamefont
  {Haakh}}, \bibinfo {author} {\bibfnamefont {S.}~\bibnamefont {Faez}}, \ and\
  \bibinfo {author} {\bibfnamefont {V.}~\bibnamefont {Sandoghdar}},\ }\href
  {\doibase 10.1103/PhysRevA.94.053840} {\bibfield  {journal} {\bibinfo
  {journal} {Phys. Rev. A}\ }\textbf {\bibinfo {volume} {94}},\ \bibinfo
  {pages} {053840} (\bibinfo {year} {2016})}\BibitemShut {NoStop}%
\bibitem [{\citenamefont {Tsoi}\ and\ \citenamefont {Law}(2008)}]{Tsoi:2008aa}%
  \BibitemOpen
  \bibfield  {author} {\bibinfo {author} {\bibfnamefont {T.~S.}\ \bibnamefont
  {Tsoi}}\ and\ \bibinfo {author} {\bibfnamefont {C.~K.}\ \bibnamefont {Law}},\
  }\href {\doibase 10.1103/PhysRevA.78.063832} {\bibfield  {journal} {\bibinfo
  {journal} {Phys. Rev. A}\ }\textbf {\bibinfo {volume} {78}},\ \bibinfo
  {pages} {063832} (\bibinfo {year} {2008})}\BibitemShut {NoStop}%
\bibitem [{\citenamefont {Asenjo-Garcia}\ \emph {et~al.}(2017)\citenamefont
  {Asenjo-Garcia}, \citenamefont {Moreno-Cardoner}, \citenamefont {Albrecht},
  \citenamefont {Kimble},\ and\ \citenamefont {Chang}}]{Asenjo-Garcia2017}%
  \BibitemOpen
  \bibfield  {author} {\bibinfo {author} {\bibfnamefont {A.}~\bibnamefont
  {Asenjo-Garcia}}, \bibinfo {author} {\bibfnamefont {M.}~\bibnamefont
  {Moreno-Cardoner}}, \bibinfo {author} {\bibfnamefont {A.}~\bibnamefont
  {Albrecht}}, \bibinfo {author} {\bibfnamefont {H.~J.}\ \bibnamefont
  {Kimble}}, \ and\ \bibinfo {author} {\bibfnamefont {D.~E.}\ \bibnamefont
  {Chang}},\ }\href {\doibase 10.1103/PhysRevX.7.031024} {\bibfield  {journal}
  {\bibinfo  {journal} {Phys. Rev. X}\ }\textbf {\bibinfo {volume} {7}},\
  \bibinfo {pages} {031024} (\bibinfo {year} {2017})}\BibitemShut {NoStop}%
\bibitem [{\citenamefont {Albrecht}\ \emph {et~al.}(2019)\citenamefont
  {Albrecht}, \citenamefont {Henriet}, \citenamefont {Asenjo-Garcia},
  \citenamefont {Dieterle}, \citenamefont {Painter},\ and\ \citenamefont
  {Chang}}]{Albrecht2018}%
  \BibitemOpen
  \bibfield  {author} {\bibinfo {author} {\bibfnamefont {A.}~\bibnamefont
  {Albrecht}}, \bibinfo {author} {\bibfnamefont {L.}~\bibnamefont {Henriet}},
  \bibinfo {author} {\bibfnamefont {A.}~\bibnamefont {Asenjo-Garcia}}, \bibinfo
  {author} {\bibfnamefont {P.~B.}\ \bibnamefont {Dieterle}}, \bibinfo {author}
  {\bibfnamefont {O.}~\bibnamefont {Painter}}, \ and\ \bibinfo {author}
  {\bibfnamefont {D.~E.}\ \bibnamefont {Chang}},\ }\href {\doibase
  10.1088/1367-2630/ab0134} {\bibfield  {journal} {\bibinfo  {journal} {New J.
  Phys.}\ }\textbf {\bibinfo {volume} {21}},\ \bibinfo {pages} {025003}
  (\bibinfo {year} {2019})}\BibitemShut {NoStop}%
\bibitem [{\citenamefont {Zhang}\ and\ \citenamefont
  {M\o{}lmer}(2019)}]{Zhang:2019aa}%
  \BibitemOpen
  \bibfield  {author} {\bibinfo {author} {\bibfnamefont {Y.-X.}\ \bibnamefont
  {Zhang}}\ and\ \bibinfo {author} {\bibfnamefont {K.}~\bibnamefont
  {M\o{}lmer}},\ }\href {\doibase 10.1103/PhysRevLett.122.203605} {\bibfield
  {journal} {\bibinfo  {journal} {Phys. Rev. Lett.}\ }\textbf {\bibinfo
  {volume} {122}},\ \bibinfo {pages} {203605} (\bibinfo {year}
  {2019})}\BibitemShut {NoStop}%
\bibitem [{\citenamefont {Yu}\ \emph {et~al.}(2020)\citenamefont {Yu},
  \citenamefont {Zhang}, \citenamefont {Sharma}, \citenamefont {Zhang},
  \citenamefont {Blanter},\ and\ \citenamefont {Bauer}}]{Yu:2020aa}%
  \BibitemOpen
  \bibfield  {author} {\bibinfo {author} {\bibfnamefont {T.}~\bibnamefont
  {Yu}}, \bibinfo {author} {\bibfnamefont {Y.-X.}\ \bibnamefont {Zhang}},
  \bibinfo {author} {\bibfnamefont {S.}~\bibnamefont {Sharma}}, \bibinfo
  {author} {\bibfnamefont {X.}~\bibnamefont {Zhang}}, \bibinfo {author}
  {\bibfnamefont {Y.~M.}\ \bibnamefont {Blanter}}, \ and\ \bibinfo {author}
  {\bibfnamefont {G.~E.~W.}\ \bibnamefont {Bauer}},\ }\href {\doibase
  10.1103/PhysRevLett.124.107202} {\bibfield  {journal} {\bibinfo  {journal}
  {Phys. Rev. Lett.}\ }\textbf {\bibinfo {volume} {124}},\ \bibinfo {pages}
  {107202} (\bibinfo {year} {2020})}\BibitemShut {NoStop}%
\bibitem [{\citenamefont {Dinc}\ \emph {et~al.}(2020)\citenamefont {Dinc},
  \citenamefont {Hayward},\ and\ \citenamefont {Bra{\'n}czyk}}]{Dinc:2020aa}%
  \BibitemOpen
  \bibfield  {author} {\bibinfo {author} {\bibfnamefont {F.}~\bibnamefont
  {Dinc}}, \bibinfo {author} {\bibfnamefont {L.~E.}\ \bibnamefont {Hayward}}, \
  and\ \bibinfo {author} {\bibfnamefont {A.~M.}\ \bibnamefont {Bra{\'n}czyk}},\
  }\href@noop {} {\enquote {\bibinfo {title} {Multi-dimensional super- and
  subradiance in waveguide quantum electrodynamics},}\ } (\bibinfo {year}
  {2020}),\ \Eprint {http://arxiv.org/abs/2003.04906} {arXiv:2003.04906
  [quant-ph]} \BibitemShut {NoStop}%
\bibitem [{\citenamefont {Brehm}\ \emph {et~al.}(2020)\citenamefont {Brehm},
  \citenamefont {Poddubny}, \citenamefont {Stehli}, \citenamefont {Wolz},
  \citenamefont {Rotzinger},\ and\ \citenamefont {Ustinov}}]{Brehm:2020aa}%
  \BibitemOpen
  \bibfield  {author} {\bibinfo {author} {\bibfnamefont {J.~D.}\ \bibnamefont
  {Brehm}}, \bibinfo {author} {\bibfnamefont {A.~N.}\ \bibnamefont {Poddubny}},
  \bibinfo {author} {\bibfnamefont {A.}~\bibnamefont {Stehli}}, \bibinfo
  {author} {\bibfnamefont {T.}~\bibnamefont {Wolz}}, \bibinfo {author}
  {\bibfnamefont {H.}~\bibnamefont {Rotzinger}}, \ and\ \bibinfo {author}
  {\bibfnamefont {A.~V.}\ \bibnamefont {Ustinov}},\ }\href@noop {} {\enquote
  {\bibinfo {title} {Waveguide bandgap engineering with an array of
  superconducting qubits},}\ } (\bibinfo {year} {2020}),\ \Eprint
  {http://arxiv.org/abs/2006.03330} {arXiv:2006.03330 [quant-ph]} \BibitemShut
  {NoStop}%
\bibitem [{\citenamefont {Kornovan}\ \emph {et~al.}(2019)\citenamefont
  {Kornovan}, \citenamefont {Corzo}, \citenamefont {Laurat},\ and\
  \citenamefont {Sheremet}}]{Kornovan:2019aa}%
  \BibitemOpen
  \bibfield  {author} {\bibinfo {author} {\bibfnamefont {D.~F.}\ \bibnamefont
  {Kornovan}}, \bibinfo {author} {\bibfnamefont {N.~V.}\ \bibnamefont {Corzo}},
  \bibinfo {author} {\bibfnamefont {J.}~\bibnamefont {Laurat}}, \ and\ \bibinfo
  {author} {\bibfnamefont {A.~S.}\ \bibnamefont {Sheremet}},\ }\href {\doibase
  10.1103/PhysRevA.100.063832} {\bibfield  {journal} {\bibinfo  {journal}
  {Phys. Rev. A}\ }\textbf {\bibinfo {volume} {100}},\ \bibinfo {pages}
  {063832} (\bibinfo {year} {2019})}\BibitemShut {NoStop}%
\bibitem [{\citenamefont {Poddubny}(2020)}]{Poddubny:2020aa}%
  \BibitemOpen
  \bibfield  {author} {\bibinfo {author} {\bibfnamefont {A.~N.}\ \bibnamefont
  {Poddubny}},\ }\href {\doibase 10.1103/PhysRevA.101.043845} {\bibfield
  {journal} {\bibinfo  {journal} {Phys. Rev. A}\ }\textbf {\bibinfo {volume}
  {101}},\ \bibinfo {pages} {043845} (\bibinfo {year} {2020})}\BibitemShut
  {NoStop}%
\bibitem [{\citenamefont {Asenjo-Garcia}\ \emph {et~al.}(2019)\citenamefont
  {Asenjo-Garcia}, \citenamefont {Kimble},\ and\ \citenamefont
  {Chang}}]{Asenjo-Garcia:2019aa}%
  \BibitemOpen
  \bibfield  {author} {\bibinfo {author} {\bibfnamefont {A.}~\bibnamefont
  {Asenjo-Garcia}}, \bibinfo {author} {\bibfnamefont {H.~J.}\ \bibnamefont
  {Kimble}}, \ and\ \bibinfo {author} {\bibfnamefont {D.~E.}\ \bibnamefont
  {Chang}},\ }\href {\doibase 10.1073/pnas.1911467116} {\bibfield  {journal}
  {\bibinfo  {journal} {Proceedings of the National Academy of Sciences}\
  }\textbf {\bibinfo {volume} {116}},\ \bibinfo {pages} {25503} (\bibinfo
  {year} {2019})}\BibitemShut {NoStop}%
\bibitem [{\citenamefont {Jenkins}\ \emph {et~al.}(2017)\citenamefont
  {Jenkins}, \citenamefont {Ruostekoski}, \citenamefont {Papasimakis},
  \citenamefont {Savo},\ and\ \citenamefont {Zheludev}}]{Jenkins2017}%
  \BibitemOpen
  \bibfield  {author} {\bibinfo {author} {\bibfnamefont {S.~D.}\ \bibnamefont
  {Jenkins}}, \bibinfo {author} {\bibfnamefont {J.}~\bibnamefont
  {Ruostekoski}}, \bibinfo {author} {\bibfnamefont {N.}~\bibnamefont
  {Papasimakis}}, \bibinfo {author} {\bibfnamefont {S.}~\bibnamefont {Savo}}, \
  and\ \bibinfo {author} {\bibfnamefont {N.~I.}\ \bibnamefont {Zheludev}},\
  }\href {\doibase 10.1103/PhysRevLett.119.053901} {\bibfield  {journal}
  {\bibinfo  {journal} {Phys. Rev. Lett.}\ }\textbf {\bibinfo {volume} {119}},\
  \bibinfo {pages} {053901} (\bibinfo {year} {2017})}\BibitemShut {NoStop}%
\bibitem [{\citenamefont {Mirhosseini}\ \emph {et~al.}(2019)\citenamefont
  {Mirhosseini}, \citenamefont {Kim}, \citenamefont {Zhang}, \citenamefont
  {Sipahigil}, \citenamefont {Dieterle}, \citenamefont {Keller}, \citenamefont
  {Asenjo-Garcia}, \citenamefont {Chang},\ and\ \citenamefont
  {Painter}}]{Mirhosseini:2019aa}%
  \BibitemOpen
  \bibfield  {author} {\bibinfo {author} {\bibfnamefont {M.}~\bibnamefont
  {Mirhosseini}}, \bibinfo {author} {\bibfnamefont {E.}~\bibnamefont {Kim}},
  \bibinfo {author} {\bibfnamefont {X.}~\bibnamefont {Zhang}}, \bibinfo
  {author} {\bibfnamefont {A.}~\bibnamefont {Sipahigil}}, \bibinfo {author}
  {\bibfnamefont {P.~B.}\ \bibnamefont {Dieterle}}, \bibinfo {author}
  {\bibfnamefont {A.~J.}\ \bibnamefont {Keller}}, \bibinfo {author}
  {\bibfnamefont {A.}~\bibnamefont {Asenjo-Garcia}}, \bibinfo {author}
  {\bibfnamefont {D.~E.}\ \bibnamefont {Chang}}, \ and\ \bibinfo {author}
  {\bibfnamefont {O.}~\bibnamefont {Painter}},\ }\href {\doibase
  10.1038/s41586-019-1196-1} {\bibfield  {journal} {\bibinfo  {journal}
  {Nature}\ }\textbf {\bibinfo {volume} {569}},\ \bibinfo {pages} {692}
  (\bibinfo {year} {2019})}\BibitemShut {NoStop}%
\bibitem [{\citenamefont {Guimond}\ \emph {et~al.}(2019)\citenamefont
  {Guimond}, \citenamefont {Grankin}, \citenamefont {Vasilyev}, \citenamefont
  {Vermersch},\ and\ \citenamefont {Zoller}}]{Guimond:2019aa}%
  \BibitemOpen
  \bibfield  {author} {\bibinfo {author} {\bibfnamefont {P.-O.}\ \bibnamefont
  {Guimond}}, \bibinfo {author} {\bibfnamefont {A.}~\bibnamefont {Grankin}},
  \bibinfo {author} {\bibfnamefont {D.~V.}\ \bibnamefont {Vasilyev}}, \bibinfo
  {author} {\bibfnamefont {B.}~\bibnamefont {Vermersch}}, \ and\ \bibinfo
  {author} {\bibfnamefont {P.}~\bibnamefont {Zoller}},\ }\href {\doibase
  10.1103/PhysRevLett.122.093601} {\bibfield  {journal} {\bibinfo  {journal}
  {Phys. Rev. Lett.}\ }\textbf {\bibinfo {volume} {122}},\ \bibinfo {pages}
  {093601} (\bibinfo {year} {2019})}\BibitemShut {NoStop}%
\bibitem [{\citenamefont {Shahmoon}\ \emph {et~al.}(2017)\citenamefont
  {Shahmoon}, \citenamefont {Wild}, \citenamefont {Lukin},\ and\ \citenamefont
  {Yelin}}]{Shahmoon:2017aa}%
  \BibitemOpen
  \bibfield  {author} {\bibinfo {author} {\bibfnamefont {E.}~\bibnamefont
  {Shahmoon}}, \bibinfo {author} {\bibfnamefont {D.~S.}\ \bibnamefont {Wild}},
  \bibinfo {author} {\bibfnamefont {M.~D.}\ \bibnamefont {Lukin}}, \ and\
  \bibinfo {author} {\bibfnamefont {S.~F.}\ \bibnamefont {Yelin}},\ }\href
  {\doibase 10.1103/PhysRevLett.118.113601} {\bibfield  {journal} {\bibinfo
  {journal} {Phys. Rev. Lett.}\ }\textbf {\bibinfo {volume} {118}},\ \bibinfo
  {pages} {113601} (\bibinfo {year} {2017})}\BibitemShut {NoStop}%
\bibitem [{\citenamefont {Rui}\ \emph {et~al.}(2020)\citenamefont {Rui},
  \citenamefont {Wei}, \citenamefont {Rubio-Abadal}, \citenamefont {Hollerith},
  \citenamefont {Zeiher}, \citenamefont {Stamper-Kurn}, \citenamefont {Gross},\
  and\ \citenamefont {Bloch}}]{Rui:2020aa}%
  \BibitemOpen
  \bibfield  {author} {\bibinfo {author} {\bibfnamefont {J.}~\bibnamefont
  {Rui}}, \bibinfo {author} {\bibfnamefont {D.}~\bibnamefont {Wei}}, \bibinfo
  {author} {\bibfnamefont {A.}~\bibnamefont {Rubio-Abadal}}, \bibinfo {author}
  {\bibfnamefont {S.}~\bibnamefont {Hollerith}}, \bibinfo {author}
  {\bibfnamefont {J.}~\bibnamefont {Zeiher}}, \bibinfo {author} {\bibfnamefont
  {D.~M.}\ \bibnamefont {Stamper-Kurn}}, \bibinfo {author} {\bibfnamefont
  {C.}~\bibnamefont {Gross}}, \ and\ \bibinfo {author} {\bibfnamefont
  {I.}~\bibnamefont {Bloch}},\ }\href@noop {} {\enquote {\bibinfo {title} {A
  subradiant optical mirror formed by a single structured atomic layer},}\ }
  (\bibinfo {year} {2020}),\ \Eprint {http://arxiv.org/abs/2001.00795}
  {arXiv:2001.00795 [quant-ph]} \BibitemShut {NoStop}%
\bibitem [{\citenamefont {Bekenstein}\ \emph {et~al.}(2020)\citenamefont
  {Bekenstein}, \citenamefont {Pikovski}, \citenamefont {Pichler},
  \citenamefont {Shahmoon}, \citenamefont {Yelin},\ and\ \citenamefont
  {Lukin}}]{Bekenstein:2020aa}%
  \BibitemOpen
  \bibfield  {author} {\bibinfo {author} {\bibfnamefont {R.}~\bibnamefont
  {Bekenstein}}, \bibinfo {author} {\bibfnamefont {I.}~\bibnamefont
  {Pikovski}}, \bibinfo {author} {\bibfnamefont {H.}~\bibnamefont {Pichler}},
  \bibinfo {author} {\bibfnamefont {E.}~\bibnamefont {Shahmoon}}, \bibinfo
  {author} {\bibfnamefont {S.~F.}\ \bibnamefont {Yelin}}, \ and\ \bibinfo
  {author} {\bibfnamefont {M.~D.}\ \bibnamefont {Lukin}},\ }\href {\doibase
  10.1038/s41567-020-0845-5} {\bibfield  {journal} {\bibinfo  {journal} {Nature
  Physics}\ }\textbf {\bibinfo {volume} {16}},\ \bibinfo {pages} {676}
  (\bibinfo {year} {2020})}\BibitemShut {NoStop}%
\bibitem [{\citenamefont {Zhang}\ \emph {et~al.}(2018)\citenamefont {Zhang},
  \citenamefont {Zhang},\ and\ \citenamefont {M\o{}lmer}}]{Zhang:2018aa}%
  \BibitemOpen
  \bibfield  {author} {\bibinfo {author} {\bibfnamefont {Y.-X.}\ \bibnamefont
  {Zhang}}, \bibinfo {author} {\bibfnamefont {Y.}~\bibnamefont {Zhang}}, \ and\
  \bibinfo {author} {\bibfnamefont {K.}~\bibnamefont {M\o{}lmer}},\ }\href
  {\doibase 10.1103/PhysRevA.98.033821} {\bibfield  {journal} {\bibinfo
  {journal} {Phys. Rev. A}\ }\textbf {\bibinfo {volume} {98}},\ \bibinfo
  {pages} {033821} (\bibinfo {year} {2018})}\BibitemShut {NoStop}%
\bibitem [{\citenamefont {Ke}\ \emph {et~al.}(2019)\citenamefont {Ke},
  \citenamefont {Poshakinskiy}, \citenamefont {Lee}, \citenamefont {Kivshar},\
  and\ \citenamefont {Poddubny}}]{Ke:2019aa}%
  \BibitemOpen
  \bibfield  {author} {\bibinfo {author} {\bibfnamefont {Y.}~\bibnamefont
  {Ke}}, \bibinfo {author} {\bibfnamefont {A.~V.}\ \bibnamefont
  {Poshakinskiy}}, \bibinfo {author} {\bibfnamefont {C.}~\bibnamefont {Lee}},
  \bibinfo {author} {\bibfnamefont {Y.~S.}\ \bibnamefont {Kivshar}}, \ and\
  \bibinfo {author} {\bibfnamefont {A.~N.}\ \bibnamefont {Poddubny}},\ }\href
  {\doibase 10.1103/PhysRevLett.123.253601} {\bibfield  {journal} {\bibinfo
  {journal} {Phys. Rev. Lett.}\ }\textbf {\bibinfo {volume} {123}},\ \bibinfo
  {pages} {253601} (\bibinfo {year} {2019})}\BibitemShut {NoStop}%
\bibitem [{\citenamefont {Schilder}\ \emph {et~al.}(2020)\citenamefont
  {Schilder}, \citenamefont {Sauvan}, \citenamefont {Sortais}, \citenamefont
  {Browaeys},\ and\ \citenamefont {Greffet}}]{Schilder:2020aa}%
  \BibitemOpen
  \bibfield  {author} {\bibinfo {author} {\bibfnamefont {N.~J.}\ \bibnamefont
  {Schilder}}, \bibinfo {author} {\bibfnamefont {C.}~\bibnamefont {Sauvan}},
  \bibinfo {author} {\bibfnamefont {Y.~R.~P.}\ \bibnamefont {Sortais}},
  \bibinfo {author} {\bibfnamefont {A.}~\bibnamefont {Browaeys}}, \ and\
  \bibinfo {author} {\bibfnamefont {J.-J.}\ \bibnamefont {Greffet}},\ }\href
  {\doibase 10.1103/PhysRevLett.124.073403} {\bibfield  {journal} {\bibinfo
  {journal} {Phys. Rev. Lett.}\ }\textbf {\bibinfo {volume} {124}},\ \bibinfo
  {pages} {073403} (\bibinfo {year} {2020})}\BibitemShut {NoStop}%
\bibitem [{\citenamefont {Bettles}\ \emph {et~al.}(2019)\citenamefont
  {Bettles}, \citenamefont {Lee}, \citenamefont {Gardiner},\ and\ \citenamefont
  {Ruostekoski}}]{Bettles:2019aa}%
  \BibitemOpen
  \bibfield  {author} {\bibinfo {author} {\bibfnamefont {R.~J.}\ \bibnamefont
  {Bettles}}, \bibinfo {author} {\bibfnamefont {M.~D.}\ \bibnamefont {Lee}},
  \bibinfo {author} {\bibfnamefont {S.~A.}\ \bibnamefont {Gardiner}}, \ and\
  \bibinfo {author} {\bibfnamefont {J.}~\bibnamefont {Ruostekoski}},\
  }\href@noop {} {\enquote {\bibinfo {title} {Quantum and nonlinear effects in
  light transmitted through planar atomic arrays},}\ } (\bibinfo {year}
  {2019}),\ \Eprint {http://arxiv.org/abs/1907.07030} {arXiv:1907.07030
  [quant-ph]} \BibitemShut {NoStop}%
\bibitem [{\citenamefont {Cremer}\ \emph {et~al.}(2020)\citenamefont {Cremer},
  \citenamefont {Plankensteiner}, \citenamefont {Moreno-Cardoner},
  \citenamefont {Ostermann},\ and\ \citenamefont {Ritsch}}]{Cremer:2020aa}%
  \BibitemOpen
  \bibfield  {author} {\bibinfo {author} {\bibfnamefont {J.}~\bibnamefont
  {Cremer}}, \bibinfo {author} {\bibfnamefont {D.}~\bibnamefont
  {Plankensteiner}}, \bibinfo {author} {\bibfnamefont {M.}~\bibnamefont
  {Moreno-Cardoner}}, \bibinfo {author} {\bibfnamefont {L.}~\bibnamefont
  {Ostermann}}, \ and\ \bibinfo {author} {\bibfnamefont {H.}~\bibnamefont
  {Ritsch}},\ }\href@noop {} {\enquote {\bibinfo {title} {Polarization control
  of radiation and energy flow in dipole-coupled nanorings},}\ } (\bibinfo
  {year} {2020}),\ \Eprint {http://arxiv.org/abs/2004.09861} {arXiv:2004.09861
  [quant-ph]} \BibitemShut {NoStop}%
\bibitem [{\citenamefont {Dung}\ \emph {et~al.}(2002)\citenamefont {Dung},
  \citenamefont {Kn\"oll},\ and\ \citenamefont {Welsch}}]{Dung2002}%
  \BibitemOpen
  \bibfield  {author} {\bibinfo {author} {\bibfnamefont {H.~T.}\ \bibnamefont
  {Dung}}, \bibinfo {author} {\bibfnamefont {L.}~\bibnamefont {Kn\"oll}}, \
  and\ \bibinfo {author} {\bibfnamefont {D.-G.}\ \bibnamefont {Welsch}},\
  }\href {\doibase 10.1103/PhysRevA.66.063810} {\bibfield  {journal} {\bibinfo
  {journal} {Phys. Rev. A}\ }\textbf {\bibinfo {volume} {66}},\ \bibinfo
  {pages} {063810} (\bibinfo {year} {2002})}\BibitemShut {NoStop}%
\bibitem [{\citenamefont {Alase}\ \emph {et~al.}(2016)\citenamefont {Alase},
  \citenamefont {Cobanera}, \citenamefont {Ortiz},\ and\ \citenamefont
  {Viola}}]{Alase:2016aa}%
  \BibitemOpen
  \bibfield  {author} {\bibinfo {author} {\bibfnamefont {A.}~\bibnamefont
  {Alase}}, \bibinfo {author} {\bibfnamefont {E.}~\bibnamefont {Cobanera}},
  \bibinfo {author} {\bibfnamefont {G.}~\bibnamefont {Ortiz}}, \ and\ \bibinfo
  {author} {\bibfnamefont {L.}~\bibnamefont {Viola}},\ }\href {\doibase
  10.1103/PhysRevLett.117.076804} {\bibfield  {journal} {\bibinfo  {journal}
  {Phys. Rev. Lett.}\ }\textbf {\bibinfo {volume} {117}},\ \bibinfo {pages}
  {076804} (\bibinfo {year} {2016})}\BibitemShut {NoStop}%
\bibitem [{\citenamefont {Cobanera}\ \emph {et~al.}(2017)\citenamefont
  {Cobanera}, \citenamefont {Alase}, \citenamefont {Ortiz},\ and\ \citenamefont
  {Viola}}]{Cobanera:2017aa}%
  \BibitemOpen
  \bibfield  {author} {\bibinfo {author} {\bibfnamefont {E.}~\bibnamefont
  {Cobanera}}, \bibinfo {author} {\bibfnamefont {A.}~\bibnamefont {Alase}},
  \bibinfo {author} {\bibfnamefont {G.}~\bibnamefont {Ortiz}}, \ and\ \bibinfo
  {author} {\bibfnamefont {L.}~\bibnamefont {Viola}},\ }\href {\doibase
  10.1088/1751-8121/aa6046} {\bibfield  {journal} {\bibinfo  {journal} {Journal
  of Physics A: Mathematical and Theoretical}\ }\textbf {\bibinfo {volume}
  {50}},\ \bibinfo {pages} {195204} (\bibinfo {year} {2017})}\BibitemShut
  {NoStop}%
\bibitem [{\citenamefont {Alase}\ \emph {et~al.}(2017)\citenamefont {Alase},
  \citenamefont {Cobanera}, \citenamefont {Ortiz},\ and\ \citenamefont
  {Viola}}]{Alase:2017aa}%
  \BibitemOpen
  \bibfield  {author} {\bibinfo {author} {\bibfnamefont {A.}~\bibnamefont
  {Alase}}, \bibinfo {author} {\bibfnamefont {E.}~\bibnamefont {Cobanera}},
  \bibinfo {author} {\bibfnamefont {G.}~\bibnamefont {Ortiz}}, \ and\ \bibinfo
  {author} {\bibfnamefont {L.}~\bibnamefont {Viola}},\ }\href {\doibase
  10.1103/PhysRevB.96.195133} {\bibfield  {journal} {\bibinfo  {journal} {Phys.
  Rev. B}\ }\textbf {\bibinfo {volume} {96}},\ \bibinfo {pages} {195133}
  (\bibinfo {year} {2017})}\BibitemShut {NoStop}%
\bibitem [{sp()}]{sp}%
  \BibitemOpen
  \href@noop {} {\emph {\bibinfo {title} {Supplemental Material}}}\BibitemShut
  {NoStop}%
\bibitem [{\citenamefont {Su}\ \emph {et~al.}(1979)\citenamefont {Su},
  \citenamefont {Schrieffer},\ and\ \citenamefont {Heeger}}]{Su:1979aa}%
  \BibitemOpen
  \bibfield  {author} {\bibinfo {author} {\bibfnamefont {W.~P.}\ \bibnamefont
  {Su}}, \bibinfo {author} {\bibfnamefont {J.~R.}\ \bibnamefont {Schrieffer}},
  \ and\ \bibinfo {author} {\bibfnamefont {A.~J.}\ \bibnamefont {Heeger}},\
  }\href {\doibase 10.1103/PhysRevLett.42.1698} {\bibfield  {journal} {\bibinfo
   {journal} {Phys. Rev. Lett.}\ }\textbf {\bibinfo {volume} {42}},\ \bibinfo
  {pages} {1698} (\bibinfo {year} {1979})}\BibitemShut {NoStop}%
\bibitem [{\citenamefont {Cooper}\ \emph {et~al.}(2019)\citenamefont {Cooper},
  \citenamefont {Dalibard},\ and\ \citenamefont {Spielman}}]{Cooper:2019aa}%
  \BibitemOpen
  \bibfield  {author} {\bibinfo {author} {\bibfnamefont {N.~R.}\ \bibnamefont
  {Cooper}}, \bibinfo {author} {\bibfnamefont {J.}~\bibnamefont {Dalibard}}, \
  and\ \bibinfo {author} {\bibfnamefont {I.~B.}\ \bibnamefont {Spielman}},\
  }\href {\doibase 10.1103/RevModPhys.91.015005} {\bibfield  {journal}
  {\bibinfo  {journal} {Rev. Mod. Phys.}\ }\textbf {\bibinfo {volume} {91}},\
  \bibinfo {pages} {015005} (\bibinfo {year} {2019})}\BibitemShut {NoStop}%
\bibitem [{\citenamefont {Wang}\ and\ \citenamefont
  {Zhao}(2018)}]{Wang:2018aa}%
  \BibitemOpen
  \bibfield  {author} {\bibinfo {author} {\bibfnamefont {B.~X.}\ \bibnamefont
  {Wang}}\ and\ \bibinfo {author} {\bibfnamefont {C.~Y.}\ \bibnamefont
  {Zhao}},\ }\href {\doibase 10.1103/PhysRevA.98.023808} {\bibfield  {journal}
  {\bibinfo  {journal} {Phys. Rev. A}\ }\textbf {\bibinfo {volume} {98}},\
  \bibinfo {pages} {023808} (\bibinfo {year} {2018})}\BibitemShut {NoStop}%
\bibitem [{\citenamefont {Pocock}\ \emph {et~al.}(2018)\citenamefont {Pocock},
  \citenamefont {Xiao}, \citenamefont {Huidobro},\ and\ \citenamefont
  {Giannini}}]{Pocock:2018aa}%
  \BibitemOpen
  \bibfield  {author} {\bibinfo {author} {\bibfnamefont {S.~R.}\ \bibnamefont
  {Pocock}}, \bibinfo {author} {\bibfnamefont {X.}~\bibnamefont {Xiao}},
  \bibinfo {author} {\bibfnamefont {P.~A.}\ \bibnamefont {Huidobro}}, \ and\
  \bibinfo {author} {\bibfnamefont {V.}~\bibnamefont {Giannini}},\ }\href
  {\doibase 10.1021/acsphotonics.8b00117} {\bibfield  {journal} {\bibinfo
  {journal} {ACS Photonics}\ }\textbf {\bibinfo {volume} {5}},\ \bibinfo
  {pages} {2271} (\bibinfo {year} {2018})}\BibitemShut {NoStop}%
\bibitem [{\citenamefont {Atala}\ \emph {et~al.}(2013)\citenamefont {Atala},
  \citenamefont {Aidelsburger}, \citenamefont {Barreiro}, \citenamefont
  {Abanin}, \citenamefont {Kitagawa}, \citenamefont {Demler},\ and\
  \citenamefont {Bloch}}]{Atala:2013aa}%
  \BibitemOpen
  \bibfield  {author} {\bibinfo {author} {\bibfnamefont {M.}~\bibnamefont
  {Atala}}, \bibinfo {author} {\bibfnamefont {M.}~\bibnamefont {Aidelsburger}},
  \bibinfo {author} {\bibfnamefont {J.~T.}\ \bibnamefont {Barreiro}}, \bibinfo
  {author} {\bibfnamefont {D.}~\bibnamefont {Abanin}}, \bibinfo {author}
  {\bibfnamefont {T.}~\bibnamefont {Kitagawa}}, \bibinfo {author}
  {\bibfnamefont {E.}~\bibnamefont {Demler}}, \ and\ \bibinfo {author}
  {\bibfnamefont {I.}~\bibnamefont {Bloch}},\ }\href {\doibase
  10.1038/nphys2790} {\bibfield  {journal} {\bibinfo  {journal} {Nature
  Physics}\ }\textbf {\bibinfo {volume} {9}},\ \bibinfo {pages} {795} (\bibinfo
  {year} {2013})}\BibitemShut {NoStop}%
\bibitem [{\citenamefont {Chang}\ \emph {et~al.}(2012)\citenamefont {Chang},
  \citenamefont {Jiang}, \citenamefont {Gorshkov},\ and\ \citenamefont
  {Kimble}}]{Chang2012}%
  \BibitemOpen
  \bibfield  {author} {\bibinfo {author} {\bibfnamefont {D.~E.}\ \bibnamefont
  {Chang}}, \bibinfo {author} {\bibfnamefont {L.}~\bibnamefont {Jiang}},
  \bibinfo {author} {\bibfnamefont {A.~V.}\ \bibnamefont {Gorshkov}}, \ and\
  \bibinfo {author} {\bibfnamefont {H.~J.}\ \bibnamefont {Kimble}},\ }\href
  {\doibase https://doi.org/10.1088/1367-2630/14/6/063003} {\bibfield
  {journal} {\bibinfo  {journal} {New J. Phys.}\ }\textbf {\bibinfo {volume}
  {14}},\ \bibinfo {pages} {063003} (\bibinfo {year} {2012})}\BibitemShut
  {NoStop}%
\bibitem [{\citenamefont {Leykam}\ and\ \citenamefont
  {Flach}(2018)}]{Leykam:2018aa}%
  \BibitemOpen
  \bibfield  {author} {\bibinfo {author} {\bibfnamefont {D.}~\bibnamefont
  {Leykam}}\ and\ \bibinfo {author} {\bibfnamefont {S.}~\bibnamefont {Flach}},\
  }\href {\doibase 10.1063/1.5034365} {\bibfield  {journal} {\bibinfo
  {journal} {APL Photonics}\ }\textbf {\bibinfo {volume} {3}},\ \bibinfo
  {pages} {070901} (\bibinfo {year} {2018})},\ \Eprint
  {http://arxiv.org/abs/https://doi.org/10.1063/1.5034365}
  {https://doi.org/10.1063/1.5034365} \BibitemShut {NoStop}%
\bibitem [{\citenamefont {Perczel}\ \emph {et~al.}(2020)\citenamefont
  {Perczel}, \citenamefont {Borregaard}, \citenamefont {Chang}, \citenamefont
  {Yelin},\ and\ \citenamefont {Lukin}}]{Perczel:2020aa}%
  \BibitemOpen
  \bibfield  {author} {\bibinfo {author} {\bibfnamefont {J.}~\bibnamefont
  {Perczel}}, \bibinfo {author} {\bibfnamefont {J.}~\bibnamefont {Borregaard}},
  \bibinfo {author} {\bibfnamefont {D.~E.}\ \bibnamefont {Chang}}, \bibinfo
  {author} {\bibfnamefont {S.~F.}\ \bibnamefont {Yelin}}, \ and\ \bibinfo
  {author} {\bibfnamefont {M.~D.}\ \bibnamefont {Lukin}},\ }\href {\doibase
  10.1103/PhysRevLett.124.083603} {\bibfield  {journal} {\bibinfo  {journal}
  {Phys. Rev. Lett.}\ }\textbf {\bibinfo {volume} {124}},\ \bibinfo {pages}
  {083603} (\bibinfo {year} {2020})}\BibitemShut {NoStop}%
\bibitem [{\citenamefont {Parameswaran}\ \emph {et~al.}(2013)\citenamefont
  {Parameswaran}, \citenamefont {Roy},\ and\ \citenamefont
  {Sondhi}}]{Parameswaran:2013aa}%
  \BibitemOpen
  \bibfield  {author} {\bibinfo {author} {\bibfnamefont {S.~A.}\ \bibnamefont
  {Parameswaran}}, \bibinfo {author} {\bibfnamefont {R.}~\bibnamefont {Roy}}, \
  and\ \bibinfo {author} {\bibfnamefont {S.~L.}\ \bibnamefont {Sondhi}},\
  }\href {\doibase 10.1016/j.crhy.2013.04.003} {\bibfield  {journal} {\bibinfo
  {journal} {Comptes Rendus Physique}\ }\textbf {\bibinfo {volume} {14}},\
  \bibinfo {pages} {816} (\bibinfo {year} {2013})}\BibitemShut {NoStop}%
\bibitem [{\citenamefont {Mukherjee}\ \emph {et~al.}(2015)\citenamefont
  {Mukherjee}, \citenamefont {Spracklen}, \citenamefont {Choudhury},
  \citenamefont {Goldman}, \citenamefont {\"Ohberg}, \citenamefont
  {Andersson},\ and\ \citenamefont {Thomson}}]{Mukherjee:2015aa}%
  \BibitemOpen
  \bibfield  {author} {\bibinfo {author} {\bibfnamefont {S.}~\bibnamefont
  {Mukherjee}}, \bibinfo {author} {\bibfnamefont {A.}~\bibnamefont
  {Spracklen}}, \bibinfo {author} {\bibfnamefont {D.}~\bibnamefont
  {Choudhury}}, \bibinfo {author} {\bibfnamefont {N.}~\bibnamefont {Goldman}},
  \bibinfo {author} {\bibfnamefont {P.}~\bibnamefont {\"Ohberg}}, \bibinfo
  {author} {\bibfnamefont {E.}~\bibnamefont {Andersson}}, \ and\ \bibinfo
  {author} {\bibfnamefont {R.~R.}\ \bibnamefont {Thomson}},\ }\href {\doibase
  10.1103/PhysRevLett.114.245504} {\bibfield  {journal} {\bibinfo  {journal}
  {Phys. Rev. Lett.}\ }\textbf {\bibinfo {volume} {114}},\ \bibinfo {pages}
  {245504} (\bibinfo {year} {2015})}\BibitemShut {NoStop}%
\bibitem [{\citenamefont {Goldman}\ \emph {et~al.}(2011)\citenamefont
  {Goldman}, \citenamefont {Urban},\ and\ \citenamefont
  {Bercioux}}]{Goldman:2011aa}%
  \BibitemOpen
  \bibfield  {author} {\bibinfo {author} {\bibfnamefont {N.}~\bibnamefont
  {Goldman}}, \bibinfo {author} {\bibfnamefont {D.~F.}\ \bibnamefont {Urban}},
  \ and\ \bibinfo {author} {\bibfnamefont {D.}~\bibnamefont {Bercioux}},\
  }\href {\doibase 10.1103/PhysRevA.83.063601} {\bibfield  {journal} {\bibinfo
  {journal} {Phys. Rev. A}\ }\textbf {\bibinfo {volume} {83}},\ \bibinfo
  {pages} {063601} (\bibinfo {year} {2011})}\BibitemShut {NoStop}%
\bibitem [{\citenamefont {Leykam}\ \emph {et~al.}(2018)\citenamefont {Leykam},
  \citenamefont {Andreanov},\ and\ \citenamefont {Flach}}]{Leykam:2018ab}%
  \BibitemOpen
  \bibfield  {author} {\bibinfo {author} {\bibfnamefont {D.}~\bibnamefont
  {Leykam}}, \bibinfo {author} {\bibfnamefont {A.}~\bibnamefont {Andreanov}}, \
  and\ \bibinfo {author} {\bibfnamefont {S.}~\bibnamefont {Flach}},\ }\href
  {\doibase 10.1080/23746149.2018.1473052} {\bibfield  {journal} {\bibinfo
  {journal} {Advances in Physics: X}\ }\textbf {\bibinfo {volume} {3}},\
  \bibinfo {pages} {1473052} (\bibinfo {year} {2018})}\BibitemShut {NoStop}%
\end{thebibliography}%


\begin{thebibliography}{2}%
\makeatletter
\providecommand \@ifxundefined [1]{%
 \@ifx{#1\undefined}
}%
\providecommand \@ifnum [1]{%
 \ifnum #1\expandafter \@firstoftwo
 \else \expandafter \@secondoftwo
 \fi
}%
\providecommand \@ifx [1]{%
 \ifx #1\expandafter \@firstoftwo
 \else \expandafter \@secondoftwo
 \fi
}%
\providecommand \natexlab [1]{#1}%
\providecommand \enquote  [1]{``#1''}%
\providecommand \bibnamefont  [1]{#1}%
\providecommand \bibfnamefont [1]{#1}%
\providecommand \citenamefont [1]{#1}%
\providecommand \href@noop [0]{\@secondoftwo}%
\providecommand \href [0]{\begingroup \@sanitize@url \@href}%
\providecommand \@href[1]{\@@startlink{#1}\@@href}%
\providecommand \@@href[1]{\endgroup#1\@@endlink}%
\providecommand \@sanitize@url [0]{\catcode `\\12\catcode `\$12\catcode
  `\&12\catcode `\#12\catcode `\^12\catcode `\_12\catcode `\%12\relax}%
\providecommand \@@startlink[1]{}%
\providecommand \@@endlink[0]{}%
\providecommand \url  [0]{\begingroup\@sanitize@url \@url }%
\providecommand \@url [1]{\endgroup\@href {#1}{\urlprefix }}%
\providecommand \urlprefix  [0]{URL }%
\providecommand \Eprint [0]{\href }%
\providecommand \doibase [0]{http://dx.doi.org/}%
\providecommand \selectlanguage [0]{\@gobble}%
\providecommand \bibinfo  [0]{\@secondoftwo}%
\providecommand \bibfield  [0]{\@secondoftwo}%
\providecommand \translation [1]{[#1]}%
\providecommand \BibitemOpen [0]{}%
\providecommand \bibitemStop [0]{}%
\providecommand \bibitemNoStop [0]{.\EOS\space}%
\providecommand \EOS [0]{\spacefactor3000\relax}%
\providecommand \BibitemShut  [1]{\csname bibitem#1\endcsname}%
\let\auto@bib@innerbib\@empty
\bibitem [{\citenamefont {Asenjo-Garcia}\ \emph {et~al.}(2017)\citenamefont
  {Asenjo-Garcia}, \citenamefont {Moreno-Cardoner}, \citenamefont {Albrecht},
  \citenamefont {Kimble},\ and\ \citenamefont {Chang}}]{Asenjo-Garcia2017}%
  \BibitemOpen
  \bibfield  {author} {\bibinfo {author} {\bibfnamefont {A.}~\bibnamefont
  {Asenjo-Garcia}}, \bibinfo {author} {\bibfnamefont {M.}~\bibnamefont
  {Moreno-Cardoner}}, \bibinfo {author} {\bibfnamefont {A.}~\bibnamefont
  {Albrecht}}, \bibinfo {author} {\bibfnamefont {H.~J.}\ \bibnamefont
  {Kimble}}, \ and\ \bibinfo {author} {\bibfnamefont {D.~E.}\ \bibnamefont
  {Chang}},\ }\href {\doibase 10.1103/PhysRevX.7.031024} {\bibfield  {journal}
  {\bibinfo  {journal} {Phys. Rev. X}\ }\textbf {\bibinfo {volume} {7}},\
  \bibinfo {pages} {031024} (\bibinfo {year} {2017})}\BibitemShut {NoStop}%
\bibitem [{\citenamefont {Zhang}\ and\ \citenamefont
  {M\o{}lmer}(2019)}]{Zhang:2019aa}%
  \BibitemOpen
  \bibfield  {author} {\bibinfo {author} {\bibfnamefont {Y.-X.}\ \bibnamefont
  {Zhang}}\ and\ \bibinfo {author} {\bibfnamefont {K.}~\bibnamefont
  {M\o{}lmer}},\ }\href {\doibase 10.1103/PhysRevLett.122.203605} {\bibfield
  {journal} {\bibinfo  {journal} {Phys. Rev. Lett.}\ }\textbf {\bibinfo
  {volume} {122}},\ \bibinfo {pages} {203605} (\bibinfo {year}
  {2019})}\BibitemShut {NoStop}%
\end{thebibliography}%

\end{document}


\title{Supplemental Material to ``Subradiant emission from regular atomic arrays: universal scaling of decay rates from the generalized Bloch theorem''}
\author{Yu-Xiang Zhang}
\affiliation{Niels Bohr Institute, University of Copenhagen, Blegdamsvej 17, 2100 Copenhagen, Denmark}
\author{Klaus M{\o}lmer}
\affiliation{Center for Complex Quantum Systems, Department of Physics and Astronomy, Aarhus University, 8000 Aarhus C, Denmark}
\date{\today}
\maketitle

This Supplemental Material is organized as follows: 
In Sec.~\ref{DDT-a4}, we show that the band of singly excited states of an array of emitters has power law dispersion relation with  $s=4$ when $k_0=k_{(4)}$; 
In Sec.~\ref{ImH}, we show that periodic Hamiltonians, restricted to a finite lattice can be written in the form of Eq.~(3) of the main text;
In Sec.~\ref{B-M}, we derive Eq.~(8) of the main text;
In Sec.~\ref{rehs}, we elaborate the proof about the 
perturbative treatment of $\Delta H=H_{\eff}^{\text{Re}}-\mathbf{H}_{s/2}$
with respect to the eigenstates of  $\mathbf{H}_{s/2}$ with wavenumbers near $k_\ex$;
In Sec.~\ref{toy-model}, we derive
analytical expressions for the eigenstates of $\mathbf{H}_1$, $\mathbf{H}_2$, and the toy model $H_{R=2}$ with
$h_1> 4h_2$ and $h_1 < 4h_2$ (the case where the extremum is degenerate). 
The results confirm Eq.~(13) of the main text.
In Sec.~\ref{appendix-SSH}, we show the mapping between $H_{\kd}^{-1}$ and the original SSH model, and we derive Eq.~(16) of the main text.

\section{The dyadic Green's tensor and the power $s$ in the dispersion relation near $k_\ex=\pi/d$}\label{DDT-a4}
The dyadic Green's tensor for the electromagnetic field in the 3D free space in vacuum is given by
\begin{equation}
\begin{aligned}
\mathbf{G}_0(\mathbf{r},\omega_0)= & \frac{e^{ik_0 r}}{4\pi k_0^3 r^3}\bigg[ (k^2_0r^2+ ik_0r-1)\mathbb{I}+\\
& (-k^2_0 r^2-3ik_0r+3)\frac{\mathbf{rr}}{r^2}\bigg],
\end{aligned}
\end{equation}
where $\mathbb{I}$ is the $3\times3$ identity tensor and $r$ is the magnitude of $\mathbf{r}$.
The energy shift for the array of emitters reads ~\cite{Asenjo-Garcia2017}
\begin{equation*}
\omega_k=\frac{3\gamma_0}{4}
\sum_{\varepsilon=\pm 1}\sum_{\xi=1}^{3} i(\frac{i}{k_0 d})^{\xi}\mathrm{Li}_\xi [ e^{i(k_0+\varepsilon k)d}],
\end{equation*}
where $\mathrm{Li}_{\xi}(z)=\sum_{n=1}^{\infty}z^n n^{-\xi}$ is the polylogarithm of order $\xi$.
The derivatives of $\omega_k$ can be evaluated by using the formula 
\begin{equation*}
\frac{d}{dz}\mathrm{Li}_{\xi}(z)=\frac{1}{z}\mathrm{Li}_{\xi-1}(z).
\end{equation*}
Although $\mathrm{Li}_{\xi}$ with $\xi<1$ are not well-defined by the series, they can
be defined through the above formula and
the equality $\mathrm{Li}_{1}(z)=-\mathrm{ln}(1-z)$. 

The parity symmetry ($\omega_k=\omega_{-k}$) makes all the odd order derivatives of 
$\omega_k$ vanish at $k=\pi/d$.
The second order derivative of $\omega_k$ at $k=\pi/d$ is
\begin{equation*}
\begin{aligned}
\partial_{k}^2\omega_{k}\bigg\rvert _{\pi/d}=& \frac{3d}{2 k_0 } \bigg[
\ln(2\cos\frac{k_0 d}{2})+\frac{k_0 d}{2}\tan(\frac{k_0 d}{2}) \\
&\qquad -(\frac{k_0 d}{2})^2\frac{1}{\cos^2(\frac{k_0 d}{2})}  \bigg].
\end{aligned}
\end{equation*}
A numerical search shows that the second derivative vanishes at $k_0=k_{(4)}\approx 0.48280076\pi/d$.

\section{Derivation of Eq.~(3) of the main text}
\label{ImH}
In this section, a representation of
discrete-translation-invariant Hamiltonians of finite chains will be introduced, which
formalizes $H^{\text{Im}}_{\eff}$ into the form of Eq.~(3) of the main text. 
The representation is based on Bloch states restricted to a finite lattice, and we shall start with a brief review of the infinite lattice limit.

\subsection{a brief review of the Bloch states of infinite lattice}
It is well-known that on infinite periodic lattices, Hamiltonians which preserve the
discrete translation symmetry are diagonalized by the Bloch states
\begin{equation}\label{2-km}
    \ket{k}=\sum_{m=-\infty}^{\infty} e^{ik x_m} \ket{m}.
\end{equation}
The Bloch states are normalized to the Dirac $\delta$-functions $\braket{k|k'}=\frac{2\pi}{d}\delta(k-k')$, with values of $k$ restricted to
the Brillouin Zone $[-\pi/d, \pi/d]$. The Bloch states constitute
an orthonormal basis. The transformation in
Eq.~\eqref{2-km} can be inverted as
\begin{equation}\label{2-mk}
    \ket{m}=\int_{BZ}\frac{\mathrm{d}k}{2\pi/d}\,e^{-ik x_m}\ket{k},
\end{equation}
where the subscript ``BZ'' means that the integral is over the Brillouin Zone $k \in [-\pi/d,\pi/d]$. One can readily check that the
above transformation is consistent with the normalization 
$\braket{m |n}=\delta_{m,n}$.

To see how the Bloch states diagonalize the Hamiltonian, we 
consider a Hamiltonian generally written as
\begin{equation}\label{2-H}
H^{\infty}=\sum_{m,n=-\infty}^{\infty} h_{m,n}\ket{m}\bra{n},
\end{equation}
which preserves the discrete translation symmetry in the sense that
$h_{m,m+r}=h(r)$. By substituting Eq.~\eqref{2-mk} into Eq.~\eqref{2-H} we obtain that
\begin{equation}
\begin{aligned}
H^{\infty} =&(\frac{1}{2\pi/d})^2\int_{BZ}\mathrm{d}k
\int_{BZ}\mathrm{d}k'\sum_{m=-\infty}^{\infty}
    \sum_{r} h_{m,m+r} \\
    &\times e^{i(k'-k)x_m}e^{ik'rd}\ket{k}\bra{k'}.
\end{aligned}
\end{equation}
Using the quasi-translation invariance $h_{m,m+r}=h(r)$, the summation of $m$ and $r$ can be implemented separately. Then by using the relation
\begin{equation}
\sum_{m=-\infty}^{\infty}e^{i(k'-k)x_m}=\frac{2\pi}{d}\delta(k'-k)
\end{equation}
and the dispersion relation $\omega_k=\sum_{r}h(r)e^{irkd}$,
we obtain the diagonal form
\begin{equation}\label{2-diagonal}
H^{\infty}=\int_{BZ}\frac{\mathrm{d}k}{2\pi/d}\;\omega_k \ket{k}\bra{k}.
\end{equation}

\subsection{the reduction to finite lattices}
A finite lattice with open boundary conditions is obtained by
truncating Eq.~\eqref{2-H} to $N$ neighboring sites. The
truncated Hamiltonian, denoted by $H_{(N)}$,
is hence obtained from $H^{\infty}$ by keeping only the lattices sites $1,2,\cdots N$.
The reduction of the Hilbert space can be indicated by a projector $P_{N}$ so that 
$$H_{(N)}=P_N H^{\infty} P_N=\sum_{m,n=1}^{N} h_{m,n}\ket{m}\bra{n}.$$
Meanwhile, from Eq.~\eqref{2-diagonal} we get that
\begin{equation}\label{2-Pbloch}
H_{(N)}=\int_{BZ}\frac{\mathrm{d}k}{2\pi/d}\, \omega_k\; P_{N}\ket{k}\bra{k}P_{N}
\end{equation}
where $P_{N}\ket{k}=\sum_{m=1}^{N} e^{ik x_m}\ket{m}$.
We caution against the potential misconception that $H_{(N)}$ is therefore diagonalized by $P_{N}\ket{k}$.This is not the case as the reduced Bloch states are not
orthogonal.

$P_{N}\ket{k}$ is not normalized to the Dirac $\delta$-function, but we have $\braket{k| P_{N} |k}=N$. This suggest defining the Bloch states on the finite lattice as
\begin{equation}
    \ket{z=e^{ikd}}\equiv\frac{1}{\sqrt{N}}P_{N}\ket{k}=
    \frac{1}{\sqrt{N}}\sum_{m=1}^{N}e^{ik x_m}\ket{m}
\end{equation}
so that $\braket{z|z}=1$. In terms of $\ket{z}$, the Hamiltonian of the finite lattice
can thus be represented as
\begin{equation}
H_{(N)} = N \int_{BZ} \frac{dk}{2\pi/d} \,\omega_k \ket{e^{ikd}}\bra{e^{ikd}}.
\end{equation}

\subsection{Dipole-dipole interaction Hamiltonian}
It is shown in Ref.~\cite{Zhang:2019aa} that the effective 
Hamiltonian of the regular array of atomic emitters can be written as
\begin{equation*}
\begin{aligned}
H_\eff=&-i\frac{3\gamma_0}{4k_0}\int_{0}^{k_0}\frac{d k}{2\pi}
\rho_{+}(k)\sum_{m,n=1}^{N}e^{i k |z_m-z_n|}\sigma_m^\dagger \sigma_n\\
&-\frac{3\gamma_0}{4k_0}\int_{0}^{+\infty}\frac{d k}{2\pi} \rho_{-}(k)
\sum_{m,n=1}^{N}e^{-k |z_m-z_n|}\sigma_m^\dagger \sigma_n.
\end{aligned}
\end{equation*}
where $\rho_{\pm}(k)$ equal $\pi(1\pm k^2/{k_0^2})$ for $k\in [-k_0, k_0]$ 
and vanish otherwise. Following the notation of the main text,
we denote the interval $[-k_0, k_0]$ by $\Gamma$ so that the dissipative part of the Hamiltonian in Eq.~(4) of the main text is given as
\begin{equation*}
H_{\eff}^{\text{Im}}=N\frac{3\gamma_0}{4k_0}\int_{\Gamma}\frac{d k}{4\pi}
\rho_{+}(k)\ket{e^{ikd}}\bra{e^{ikd}}.
\end{equation*}

\section{Derivation of Eq.~(8) of the main text}\label{B-M}
While we may restrict the generalized Bloch theorem to Hermitian Hamiltonians $H_R$, it applies also to more general Hamiltonians
\begin{equation}
    H_R=h_0\mathbb{I}+\sum_{r=1}^{R}\sum_{m=1}^{N-r} h_r\ket{m}\bra{m+r} + h^{'}_r \ket{m+r}\bra{m},
\end{equation}
where $h^{'}_{r}$ is not necessarily $h_r^{*}$.
The boundary equation $P_\partial(H_R-E)\ket{\psi}=0$ can be formulated through a
boundary matrix $M$
defined with elements 
\begin{equation}
M_{b,j}\equiv\sqrt{N}\braket{b|(H_R-E)|z_j},
\end{equation}
where $\bra{b}$ denotes an excitation located at the $b$th boundary site.
Then the boundary equation is equivalent to
$\sum_j M_{b,j}c_j=0$ for any $b\in \partial_l\cup\partial_r$.

Plain calculation leads to the expression
\begin{equation}\label{M-element}
M_{b,j}=(h_0-E)z_j^b+\sum_{r=1}^{R}(\theta_{b+r}h_r z_j^{b+r}+\theta_{b-r}h'_r z_j^{b-r}),
\end{equation}
where $\theta_x=1 $ for $x\in\partial_l\cup\partial_r$ and $\theta_x=0$ otherwise.
By definition, $z_j$ satisfies $\tilde{\omega}_{R}(z_j)=E$, i.e.,
\begin{equation*}
h_0+\sum_{r=1}^{R}(h_r z_j^r+h'_r z_j^{-r})=E.
\end{equation*}
We substitute the above equality into Eq.~\eqref{M-element}
to eliminate $E$. It results in
\begin{equation*}
M_{b,j}=z_j^b\sum_{r=1}^{R}\bigg[(1-\theta_{b+r})h_r z_j^{r}+(1-\theta_{b-r})h'_r z_j^{-r}\bigg]
\end{equation*}
Explicitly, for $b\in \partial_l=\{1,2,\cdots R \}$, we have
\begin{equation}
\begin{aligned}
& M_{R,j} =h'_R, \\
& M_{R-1, j} = h'_{R} z_j^{-1} +h'_{R-1}, \\
& \cdots \\
& M_{1,j}  = h'_{R} z_j^{-(R-1)} +\cdots + h'_2 z_j^{-1} + h'_1.
\end{aligned}
\end{equation}
Formulae for $b\in\partial_r=\{ N, {N-1},\cdots, {N-R+1}\}$ 
will be given later.

Note that to find $\{ c_j\}_j$ satisfying $\sum_{j}M_{b,j}c_j=0$,
we are free to manipulate the matrix $M$ by row operations.
To proceed, the first step is to divide $M_{R,j}$ by a factor of $h'_R$:
\begin{equation*}
M_{R,j}\rightarrow M_{R,j}/h'_R=1.
\end{equation*}
Next, we subtract them from the second row:
\begin{equation*}
\begin{aligned}
M_{R-1,j} & \rightarrow (M_{R-1, j}- h'_{R-1} M_{R,j})/h'_R \\
& = z_j^{-1}.
\end{aligned}
\end{equation*}
Similar manipulations are applied for the remaining rows. Finally, we obtain
a normal form for $M_{b,j}\;(b\in\partial_l)$:
\begin{equation}\label{upper-M}
\begin{aligned}
& M_{R,j}  = 1, \\
& M_{R-1,j}  = z_j^{-1}, \\
& \cdots \\
& M_{1,j}  = z_j^{-(R-1)},
\end{aligned}
\end{equation}
and for $M_{b,j}\;(b\in\partial_r)$
\begin{equation}\label{lower-M}
\begin{aligned}
& M_{N-R+1, j}  = z_j^{N+1}, \\
& M_{N-R+2, j}  = z_j^{N+2}, \\
& \cdots \\
& M_{N,j}  = z_j^{N+R}.
\end{aligned}
\end{equation}
It is convenient to denote the sub-matrix of
$b\in\partial_l$ by $M^L$ and
that of $b\in\partial_r$ by $M^{R}$, and relabel the matrix elements as
\begin{equation}\label{relabel-M}
M^{L}_{r,j}=z_j^{-r},\quad
M^{R}_{r,j}=z_j^{N+1+r},
\end{equation}
where $r=0,1,\cdots, R-1$. The equation $\sum_j M_{b,j}c_j=0$ for every $b$
is then equivalent to
\begin{equation}\label{virtual-site}
\sum_j z_j^{-r}c_j=0,\quad \sum_{j}z_j^{N+1+r}c_j=0.
\end{equation}
for every $r\in\{0,1,\cdots, R-1\}$.

To obtain Eq.~(8) of the main text, we need to rewrite the above formulae with 
$\epsilon_j$ and $\eta_j$. 
First, we multiply the $r$th row of $M^L$ by 
$z_{\ex}^{r}$ so that
\begin{equation*}
M^{L}_{r,j} \rightarrow (1+\epsilon_j)^{r}=\sum_{k=0}^{r} C_{r}^{k}\epsilon_j^k,
\end{equation*}
where $C_{r}^{k}=\frac{r!}{k!(r-k)!}$ is the binomial coefficient. Then we sequentially apply the
row operations
\begin{equation*}
M^{L}_{r,j}\rightarrow  M^{L}_{r,j}-\sum_{m=0}^{r-1}C_{r}^{m} M^{L}_{m,j} 
\end{equation*}
by the order of $r=1,2,\cdots$, to reshape the rows into
\begin{equation}
M^L_{r,j}=\epsilon_j^{r}.
\end{equation}
Similarly, for $M^R$, we multiply $M^{R}_{r,j}$ by $z_{\ex}^{-r}$ and obtain
\begin{equation*}
M^{R}_{r,j}=z_j^{N+1}(1+\eta_j)^{r}=z_j^{N+1}\sum_{k=0}^{r}C_{r}^{k}\eta_j^k.
\end{equation*}
Then row manipulations lead to $M^R_{r,j}=z_j^{N+1}\eta_j^{r}$ so that
Eq.~(8) of the main text is obtained.

\section{The perturbative relation between $H_{\eff}^{\text{Re}}$ and $\mathbf{H}_{s/2}$}
\label{rehs}

In the section, we elaborate on the relation between $H_{\eff}^{\text{Re}}$
and $\mathbf{H}_{s/2}$  discussed only briefly in the main text.
First, $\Delta H=H_{\eff}^{\text{Re}}-\mathbf{H}_{s/2}$ is 
represented in the form derived in Sec.~\ref{ImH}, or Eq.~(3) of the main text, as
\begin{equation}
\Delta H = N\int_{BZ}\frac{\mathrm{d}k}{2\pi/d}\,\delta\omega_k 
\ket{e^{ikd}}\bra{e^{ikd}},
\end{equation}
where $\delta\omega_k=\omega_k -\omega_{s/2}(k)$. To verify that $\Delta H$
is indeed a perturbation to $\ket{\psi}=\sum_{j=1}^{s}c_j\ket{z_j}$, we use
the same approach as in the main text and evaluate 
$\braket{\psi|\Delta H |\psi}$. Here we duplicate the formula
of $\braket{z=e^{ikd} | \psi}$ in the main text for the readers' convenience
\begin{equation}\label{S-rehs}
\begin{aligned}
\braket{z| \psi  } = & \frac{1}{N}\bigg[
\frac{z_{\ex}}{z-z_{\ex}}  
\sum_{n=0}^{\infty} \frac{z^n \sum_{j=1}^{s} c_j \epsilon_j^n}{(z_{\ex}-z)^{n}} \\
 & \qquad + \frac{z^{-N}}{z-z_{\ex}}\sum_{n=0}^{\infty}
  \frac{\sum_{j=1}^{s} c_j z_j^{N+1}\eta_j^n}{(z-z_{\ex})^n}
\bigg].
\end{aligned}
\end{equation}
Here we have to deal with the situation where $k$ is within a small interval 
$O_{\ex}$ around $k_{ex}$ with $\abs{k-k_{\ex}}\sim N^{-1}$. For $k\in O_{\ex}$, the denominators of Eq.~\eqref{S-rehs} are
small and require special attention. This is unlike our discussion for $H_{\eff}^{\text{Im}}$, which excluded that situation because
$\gamma_k=0$ for $k\notin \Gamma$. 

To proceed, we note that there is another neighborhood of $k_{\ex}$, say $\Omega$, inside which $\mathbf{H}_{s/2}$  
approximates $H_{\eff}^{\text{Re}}$ well. More precisely, for $k\in \Omega$ we have
\begin{equation}
\delta\omega_k=\sum_{l=s+1}^{\infty} a_{l} (k-k_{\ex})^l.
\end{equation}
Mathematically, $\Omega$ is the region where the Taylor series of $\omega_k$ converges. The size of $\Omega$, i.e.,
the radius of convergence is given by 
\begin{equation}
\abs{k-k_{\ex}} < \lim_{l\rightarrow \infty} \abs{\frac{a_l}{a_{l+1}}}.
\end{equation}
Note that the convergence radius is a mathematical property of $\omega_k$ that
does not depend on $N$. Thus, provided a large enough $N$, we will have
$O_{\ex}\subset\Omega$. By definition the size of
$O_{\ex}$ should be $\mu(O_{\ex})\sim N^{-1}$, and therefore, in the evaluation of $\braket{\psi | \Delta H |\psi}$, 
the integral for $k\in O_{\ex}$ can be estimated as
\begin{equation}
\begin{aligned}
& N\int_{O_{\ex}}\frac{\mathrm{d}k}{2\pi/d}\delta\omega_k \braket{\psi | e^{ikd}}\braket{e^{ikd} | \psi} \\
\sim &N \times N^{-1} \times N^{-s-1} \times \frac{1}{N^2} \sum_{n=s}^{\infty}\frac{N^{-n}}{(N^{-1})^{n+2}} \\
\sim & N^{-s-1},
\end{aligned}
\end{equation}
where terms in the second line correspond one-by-one to
the terms in the first line: the overall factor $N$, the size of
$O_{\ex}$ $\mu(O_{\ex})\sim N^{-1}$, the leading order of $\delta\omega_k$,
and $\abs{\braket{z=e^{ikd}|\psi}}^2$ with $k\in O_{\ex}$.

For the integral of $k\notin O_{\ex}$, the treatment is similar to the one for $H_{\eff}^{\text{Im}}$ with similar conclusions. In the next section, the perturbative relationship
will be examined further by explicitly calculating the eigenstates of the toy models.

\section{Analytical results of toy models} 
\label{toy-model}
In the main text, we have demonstrated that near
an extremum point of the dispersion relation, eigenstates of $H_R$ can be approximated
by those of $\mathbf{H}_{s/2}$. The approximation is well described
by the first order perturbation formalism with $N^{-1}$ being the small parameter. It leads to the relation presented in Eq.~(13) of the main text.

In this section, we will verify this claim by analytically calculating
and comparing the eigenstates of $\mathbf{H}_1$ and
$\mathbf{H}_2$, and a toy model $H_{R=2}$.
The model $H_{R=2}$ is assumed to have positive tunneling strengths
$h_0=0$, $h_r=h_r'$ for $r=1,2$: 
\begin{equation}\label{H2}
H_{2}=h_1\sum_{m=1}^{N-1} \ket{m}\bra{m+1} +h_2\sum_{m=1}^{N-2}\ket{m}\bra{m+2}+h.c.
\end{equation}
The dispersion relation of this model is given by
\begin{equation*}
\omega_2(k)=2h_1\cos(k)+2h_2\cos(2k).
\end{equation*}
The point $k_\ex=\pi/d$ is an extremum point with the power law  $s=4$ if $h_1=4h_2$ (now  
$H_2$ is actually $\mathbf{H}_2$).
Otherwise, it is an extremum point with $s=2$ if $h_1 > 4h_2$; or a local extremum, degenerate with to other two Bloch modes, if $h_1<4h_2$.
These cases of the toy model correspond to the emitter array coupled via the 3D electromagnetic field for
$k_0=k_{(4)}$, $0.55\pi/d$ and $0.3\pi/d$.

The eigenstates of our toy model Hamiltonians can be derived following the generalized Bloch theorem. A few useful formulae are given here in advance.
The bulk equation for eigenstates near the extremum refers to
the equation
\begin{equation}
\tilde{\omega}_{2}(z)=h_1(z+\frac{1}{z})+h_2(z^2+\frac{1}{z^2})=E,
\end{equation}
with the eigenvalue Ansatz, $E=\omega_2(k_\ex)\pm e$, where
we take the $+(-)$ sign if $k_{\ex}$ is a minimum(maximum). 
Assuming an energy minimum and using the condition 
that $e\ll h_1$,
the equation $\tilde{\omega}_{2}(z)=E$ leads to
\begin{equation}\label{h2-z}
z+\frac{1}{z}=-\frac{h_1}{2 h_2}\pm\frac{1}{2 h_2}\sqrt{(h_1-4h_2)^2+4h_2 e}.
\end{equation}
For the boundary equation, 
we write the boundary matrix $M$ in the normal form of
Eqs.~\eqref{upper-M} and~\eqref{lower-M}. Eigenstates with eigenvalue $E$ exist if
and only if the determinant of $M$ vanishes. We have
\begin{equation}\label{detM}
\begin{aligned}
\det M\propto \sum_{j_1<j_2}\sum_{j_3<j_4}\epsilon_{j_1 j_2 j_3 j_4} &
(z_{j_2}-z_{j_1})(z_{j_4}-z_{j_3})\\
& \times (z_{j_3}z_{j_4})^{N+2}.
\end{aligned}
\end{equation}
where $\epsilon_{j_1 j_2 j_3 j_4}$ is 1(-1) for even(odd) permutations 
with respect to $(1,2,3,4)$.
The boundary equation is equivalent to Eq.~\eqref{virtual-site},
from which the coefficients $\{c_j\}$ can be obtained.

\subsection{Eigenstates of $\mathbf{H}_1$}
For convenience, we assume $k_{\ex}=\pi/d$ so that
$\mathbf{H}_1$ is found to be
\begin{equation}
\mathbf{H}_1=h_1\sum_{m=1}^{N-1} \ket{m}\bra{m+1} +h.c..
\end{equation}
This is a basic tight binding model with the eigenstates 
\begin{equation}
\ket{\psi_\xi}=\frac{1}{\sqrt{2}}(\ket{k_\xi}-\ket{-k_\xi})
\end{equation}
where $k_\xi=(1-\frac{\xi}{N+1})\frac{\pi}{d}$ and $\xi=1,2,\cdots, N$.
The corresponding eigenvalues are $\omega_1(k)=2h_1\cos(k d)$.

\subsection{Eigenstates of $\mathbf{H}_2$ for $h_1 = 4 h_2$}
$\mathbf{H}_2$ with $k_\ex=\pi/d$ is given by Eq.~\eqref{H2} with $h_1=4 h_2$.
To proceed, we suppose $\sqrt{e/h_2}=\delta$ so that $z + 1/z =-2\pm\delta$. Then
the four solutions to $\tilde{\omega}_2(z)=E$ are
\begin{equation}
\begin{aligned}
&z_1 =z_2^{-1}\approx-e^{-i\sqrt{\delta}},\\
&z_3 =z_4^{-1} \approx -e^{-\sqrt{\delta}}.
\end{aligned}
\end{equation}
Directly substituting them into Eq.~\eqref{detM} and 
evaluating $\det M=0$ yields 
\begin{equation*}
\cos[(N+2)\sqrt{\delta}]\cosh[(N+2)\sqrt{\delta}]=1,
\end{equation*}
which is solved by
\begin{equation}
\sqrt{\delta}\approx \frac{\zeta \pi}{N+2}, \;\;\zeta=1.5\, ,2.5\, ,3.5 \cdots
\end{equation}
A notable feature is that the eigenstates are labelled by half integers ($\zeta$),
in contrast to the integers $\xi$ that appear in the 
eigenstates of $\mathbf{H}_1$.

The eigenstates of $\mathbf{H}_2$ near $k_\ex$ will be expressed as
\begin{equation}
\ket{\psi_{\zeta}}=\sum_j c^{(\zeta)}_j\ket{z^{(\zeta)}_j}.
\end{equation}
To find the superposition coefficients $\{ c^{(\zeta)}_{j}\}$, we suppose
$\zeta= \xi+0.5$ and obtain by using Eq.~\eqref{virtual-site}
\begin{equation}
-\frac{c^{(\zeta)}_4}{c^{(\zeta)}_3}=\frac{e^{-\zeta\pi}+(-1)^{\xi}}{e^{\zeta \pi}-(-1)^{\xi}}.
\end{equation}
For convenience, we let $c^{(\zeta)}_3=e^{\zeta \pi}+(-1)^{\xi}$ (the state $\ket{\psi_{\zeta}}$ can be easily normalized  afterwards). Then the other two coefficients will be
\begin{equation}
\begin{aligned}
&c^{(\zeta)}_1 =(-1)^{\xi}-\sinh(\zeta\pi)+i\cosh(\zeta\pi), \\
&c^{(\zeta)}_2 =(-1)^{\xi}-\sinh(\zeta\pi)-i\cosh(\zeta\pi).
\end{aligned}
\end{equation}
We note that the four coefficients have 
comparable magnitudes and all four components ($\ket{z^{(\zeta)}_j}$)
contribute to the eigenstate.

\subsection{Eigenstates of $H_2$ for $h_1 >4 h_2>0$ }\label{example-s=2}
We approximate the square root in Eq.~\eqref{h2-z} to first order in $e$ and obtain
\begin{equation*}
\sqrt{(h_1-4 h_2)^2 + 4 h_2 e}\approx |h_1 - 4h_2|+\frac{2h_2 e}{|h_1-4h_2|}.
\end{equation*}
Supposing $h_1/h_2-4=a$ ($a>0$) and $e/(a h_2)=\delta$, we recast
Eq.~\eqref{h2-z} into two equations
\begin{equation*}
\begin{aligned}
& z + 1/z=-2+\delta, \\
& \tilde{z}+1/\tilde{z}=-(a+2)-\delta.
\end{aligned}
\end{equation*}
Each of these equation will yield two roots of $\tilde{\omega}_{2}(z)=E$. The roots $z_{1,2}$ of the first equation, satisfy $\abs{z_{1,2}}=1$ (real $k$), while 
the roots of the second equation, $\tilde{z}_{3,4}$, 
have non-unit magnitudes (complex $k$). 
Correspondingly $\ket{\tilde{z}_{3,4}}$ represent boundary states that 
are exponentially suppressed in the bulk of the lattice.

Explicitly, the solutions are
\begin{equation*}
\begin{aligned}
&z_1=1/z_2 \approx -1+i \sqrt{\delta}\approx-e^{-i\sqrt{\delta}}, \\
&\tilde{z}_3=1/\tilde{z}_4 \approx -xe^{\beta\delta},
\end{aligned}
\end{equation*}
where $\beta=\frac{2}{a+2}+\frac{2}{\sqrt{a^2+4a}}$ and $x=\frac{a}{2}+1$.
Note that $\abs{\tilde{z}_3}>1$ and $\abs{\tilde{z}_3 \tilde{z}_4}=1$.
Substituting these expressions into Eq.~\eqref{detM} leads to
\begin{equation}\label{s2r2}
\sqrt{\delta}(N+2-\frac{x+1}{x-1})\approx\xi\pi
\end{equation}
where $\xi=1,2,\cdots$. Note that
\begin{equation}
\lim_{h_1/h_2\gg 1}\sqrt{\delta}=\frac{\xi\pi}{N+1},
\end{equation}
which is exactly the eigen-solutions of $\mathbf{H}_1$. The difference between $\sqrt{\delta}$ obtained for a finite $h_1/h_2$ [Eq.~\eqref{s2r2}] and in the limit $h_1\gg h_2$ scales as $O(N^{-2})$.

For the expansion coefficients, we have
\begin{equation}
\frac{\tilde{c}_4}{\tilde{c}_3}=(-1)^{N+\xi}{\tilde{z}_3}^{N+1},
\end{equation}
which implies that $\tilde{c}_3 \ket{\tilde{z}_3}$ and $\tilde{c}_4\ket{\tilde{z}_4}$
are equal in magnitudes (note that $\ket{\tilde{z}_3}$ and $\ket{\tilde{z}_4}$ are not
normalized).
The coefficients $\{c_j\}_j$ are given by
\begin{equation}
\begin{aligned}
&\frac{c_1}{\tilde{c}_4}\approx -i\frac{1-x-i\sqrt{\delta}+\beta\delta}{2\sqrt{\delta}}, \\
&\frac{c_2}{\tilde{c}_4}\approx i\frac{1-x+i\sqrt{\delta}+\beta\delta}{2\sqrt{\delta}}.
\end{aligned}
\end{equation}
where we have neglected exponentially small contribution from $\tilde{c}_3$.
It follows that the components of $c_{1,2}\ket{z_{1,2}}$ are of $O(N)$ fold larger magnitude 
than $\tilde{c}_{3,4}\ket{\tilde{z}_{3,4}}$. This confirms
Eq.~(13) of the main text.

\subsection{ $H_2$ with $0<h_1 < 4h_2$ ($k_\ex$ yields a degenerate energy eigenvalue) }
$k_\ex=\pi/d$ is a local energy maximum and we assume
the Ansatz $E=\omega_2(k_\ex)-e$ with $e>0$, and define $4-h_1/h_2=a$ and $e/(ah_2)=\delta$. This yields
\begin{equation*}
\begin{aligned}
& z+1/z=-2+\delta, \\
& \tilde{z}+1/\tilde{z}=-2+a-\delta,
\end{aligned}
\end{equation*}
with the solutions $z_1=1/z_2\approx -e^{i\sqrt{\delta}}$, and
\begin{equation*}
\tilde{z}_3=\frac{1}{\tilde{z}_4}
=\frac{a-\delta -2}{2}+\frac{i}{2}(\sqrt{4a-a^2}+\frac{(a-2)\delta}{\sqrt{4a-a^2}}).
\end{equation*}
Here $\abs{\tilde{z}_3}=\abs{\tilde{z}_4}=1$ corresponding to real values of $k$, which are wave number the modes that are degenerate with the extremum at $k_\ex=\pi/d$.
We represent the states as $\tilde{z}_3=e^{i\theta}=1/\tilde{z}_4$, 
where $\theta=\theta_0+\delta'$, 
where $\delta'$ is a function of $\delta$ that 
has the same scale as $\delta$ and vanishes for $\delta=0$.
Then $\det M=0$ leads to
\begin{equation*}
\sqrt{\delta}=\frac{\xi\pi}{N+2}(1-\eta),
\end{equation*}
where $\xi=1,2,\cdots$ and the correction term is
\begin{equation}
\eta=\frac{\tan (\theta_0/2)}{N+2}\frac{1-(-1)^{\xi+N}\cos[\theta_0(N+2)]}
{(-1)^{\xi+N}\sin[\theta_0(N+2)]}.
\end{equation}
Comparing with Eq.~\eqref{s2r2},
we see that $\eta$ contains an overall suppression factor $1/(N+2)$ and trigonometric functions that oscillate as function of $N$.

As to the superposition coefficients, we
suppose $c_1=1$ so that $c_{2}=-e^{i\phi}\approx-1-i\phi$,
where
\begin{equation*}
\phi\approx\frac{-\eta\xi\pi(-1)^{N+\xi}\sin\theta+
\sqrt{\delta}\sin[\theta(N+2)]}{(-1)^{N+\xi}\sin\theta+\sin[\theta(N+2)]+\sin[\theta(N+1)]}.
\end{equation*}
It can be seen that $\phi \sim N^{-1}$.
Coefficients of $\ket{\tilde{z}_{3(4)}}$ are
\begin{equation*}
\tilde{c}_3\approx \frac{\sqrt{\delta}}{\sin\theta}-\frac{1+e^{-i\theta}}{2\sin\theta}\phi
\end{equation*}
and $\tilde{c}_4\approx -\tilde{c}_3+i\phi$. We see that 
$\tilde{c}_{3,4}$ also oscillate as function of $N$, while the relation
$c_{1(2)}/\tilde{c}_{3(4)}=O(N)$ generally holds.
Thus, Eq.~(13) of the main text is also satisfied.

\section{dimerized emitter chain coupled to an
ideal 1D waveguide}\label{appendix-SSH}

\subsection{Relation to the original SSH model}
In a lattice of $N$ unit cells, i.e., $2N$ emitters, the
inverse $H_{\kd}^{-1}$ of the single-excitation Hamiltonian is a 
$2N\times 2N$ matrix expressed as
\begin{equation}\label{H-SSH}
H_{\kd}^{-1}=\frac{1}{\gamma_0}
\begin{pmatrix}
a & J & 0 & 0 & 0 & 0 &\cdots \\
J & e_0 & J' & 0 & 0 & 0 &\cdots \\
0 & J' & e_0 & J & 0 & 0 &\cdots \\
0 & 0 & J & e_0 & J' & 0 &\cdots \\
\,& \,&\,&\cdots &\, &\cdots &\,
\end{pmatrix}
\end{equation}
where the parameters are
\begin{equation*}
\begin{aligned}
&a=i-\cot(k_0 d_1), \;\; J'=\frac{1}{\sin(k_0 d_2)},\\ 
&J=\frac{1}{\sin(k_0 d_1)}, \quad
e_0=-\frac{\sin(k_0 d)}{\sin(k_0 d_1)\sin(k_0 d_2)}.
\end{aligned}
\end{equation*}
$H_{\kd}^{-1}$ thus only differes from the original SSH model, 
by a constant term $\frac{e_0}{\gamma_0}\mathbb{I}$ that is irrelevant
to the eigenstates, and by additional energy level shifts and dissipation $(a-e_0)/\gamma_0$ on the emitters sitting on the lattice ends.

\subsection{Eigenstates of $H_{\kd}$}
If $N$ is infinity, the eigenstates are the Bloch states 
$\ket{k, \mathbf{u}_k}$.
The dispersion relation is expressed as
\begin{equation}\label{SSH-wk}
\begin{aligned}
  \omega_\pm (k)= & \frac{\gamma_0 /2}{\cos(k d)-\cos(k_0 d)}\bigg[ \sin(k_0 d) \\
  &\pm \sqrt{J^{-2}
   +(J')^{-2}+2(JJ')^{-1}\cos(kd)}\bigg].
\end{aligned}
\end{equation}
Note that $\omega_+$ diverges at $k=\pm k_0$ 
(we assume $\sin(k_0 d)>0$ without loss of generality), while $\omega_-$ does not:
\begin{equation}
\omega_-(\pm k_0)=-\frac{\gamma_0}{2}\frac{\sin(k_0 d_1)\sin(k_0 d_2)}{\sin(k_0 d)}.
\end{equation}
The intra-cell ``spin'' of the two bands are
\begin{equation*}
\ket{\mathbf{u}^{+}_{k}} =\frac{1}{\sqrt{2}}
\begin{pmatrix}
1 \\
e^{i\phi_k}
\end{pmatrix},\; \ket{\mathbf{u}^{-}_{k}} = \frac{1}{\sqrt{2}}
\begin{pmatrix}
1 \\
-e^{i\phi_k}
\end{pmatrix}.
\end{equation*}
where $\phi_k$ is given by
\begin{equation*}
\tan\phi_k=\frac{\sin(kd)}{\cos(kd)+J/J'}.
\end{equation*}
We see from Eq.~\eqref{SSH-wk} that the band gap will
close at $k=0$ if
$\sin(k_0 d_1)=-\sin(k_0 d_2)$,
and at $k=\pi/d$ if
$\sin(k_0 d_1)=\sin(k_0 d_2)$.
We focus here on the former case. 

At the critical point $\sin(k_0 d_1)=-\sin(k_0 d_2)$, the dispersion relation 
is given by
\begin{equation}
\begin{aligned}
  \omega(k)= & \frac{\gamma_0 /2}{\cos(k d)-\cos(k_0 d)}\\
  & \times\bigg[\sin(k_0 d)\pm 2 \sin(k_0 d_1) \sin(\frac{kd}{2})\bigg].
\end{aligned}
\end{equation}
We note that it is linear near $k=0$ unlike its having and  
extremum value with $s=2$ when the system is away from the critical point.

The intra-cell ``spin'' shows a discontinuity, e.g., for the upper band we have 
\begin{equation*}
\ket{\mathbf{u}^{+}_{k>0}}=\frac{1}{\sqrt{2}}\begin{pmatrix}
                1 \\
                ie^{i\frac{kd}{2}}
                \end{pmatrix},\;\;
\ket{\mathbf{u}^{+}_{k<0}}=\frac{1}{\sqrt{2}}\begin{pmatrix}
                1 \\
                -ie^{i\frac{kd}{2}}
                \end{pmatrix}.
\end{equation*}

In our derivation of the most subradiant states with wavenumbers $k\approx 0$, we shall use $\ket{k}$ as a short hand for 
an Ansatz $\ket{k, \mathbf{n}_k}$, and
$\ket{k_\pm}$ for $\ket{k, \mathbf{u}^{\pm}_{k}}$.
For finite chains, we have 
\begin{equation*}
\begin{aligned}
H_{\kd}\ket{e^{ikd},\mathbf{n}_k}=\sum_{\epsilon=\pm} &
\omega_{k}^{\epsilon}\braket{e^{ikd},\mathbf{u}_k^{\epsilon}|e^{ikd}, \mathbf{n}_k}
\ket{e^{ikd},\mathbf{u}_k^{\epsilon}} \\
& + g_k \ket{e^{ik_0 d}, \mathbf{u}^{+}_{k_0}}
-h_k\ket{e^{-ik_0 d}, \mathbf{u}_{-k_0}^+},
\end{aligned}
\end{equation*}
where $\epsilon=\pm $ and 
\begin{subequations}
\begin{equation*}
g_k=
\frac{1}{i\sqrt{2}}\gamma_{0}\frac{e^{i(k-k_0)x_1}}{1-e^{i(k-k_0)d}}
\braket{ \mathbf{u}^{+}_{k_0}|\mathbf{n}_k},
\end{equation*}
\begin{equation*}
h_k=
\frac{1}{i\sqrt{2}}\gamma_{0}\frac{e^{i(k+k_0)(x_N+d)}}{1-e^{i(k+k_0)d}}
\braket{ \mathbf{u}^{+}_{-k_{0}}|\mathbf{n}_k}.
\end{equation*}
\end{subequations}
Thus the eigenstates should be superpositions of 
$\ket{k, \mathbf{u}^{\epsilon}_k}$ and 
$\ket{-k, \mathbf{u}^{\epsilon}_{-k}}$ that satisfy
\begin{equation}\label{ghhg}
g_{k}h_{-k}=g_{-k}h_k.
\end{equation}

For eigenstates of the upper band,
Eq.~\eqref{ghhg} is expanded as
\begin{equation}\label{solution-fix-equation}
e^{-2iNdk}\frac{\sin^2\frac{k_0 +k}{2}d}{\sin^2\frac{k_0 -k}{2}d}=
\frac{1-\sin(k_0 d_1+\frac{kd}{2})}{1+\sin(k_0 d_1-\frac{kd}{2})}.
\end{equation}
For $k\approx 0$, we substitute the ansatz
\begin{equation}
k=\epsilon\frac{\pi}{Nd}(1+\frac{1}{N}\delta)
\end{equation}
into Eq.~\eqref{solution-fix-equation} and obtain
\begin{subequations}
\begin{equation*}
e^{-i2\pi\epsilon}=\frac{1-\sin k_0 d_1}{1+ \sin k_0 d_1},
\end{equation*}
\begin{equation*}
i\delta=\cot(\frac{k_0 d}{2})+\frac{1}{2}\tan(k_0 d_1). 
\end{equation*}
\end{subequations}
The solution is not unique and we have
\begin{equation}
\epsilon_{\xi}=\xi+i\frac{1}{2\pi}\ln\frac{1-\sin k_0 d_1}{1+\sin k_0 d_1},
\end{equation}
where $\xi=1,2,3\cdots$. 
Substituting the solutions of $k$ into $\omega_{+}(k)$,  we obtain
the frequency shift and the decay rates
\begin{subequations}
\begin{equation}
\omega_{+}(\xi)=\frac{\gamma_0 /2}{1-\cos k_0 d}(\sin k_0 d+ \xi\frac{\pi}{N}\sin k_0 d_1),
\end{equation}
\begin{equation}
\gamma_{+}=-\frac{\gamma_0}{4N}\cot(k_0 d_1)
\ln(\frac{1-\sin k_0 d_1}{1+\sin k_0 d_1}).
\end{equation}
\end{subequations}
Note that $\gamma_{+}$ is uniform for all $\xi$, and $\gamma_{+}\propto N^{-1}$.

For eigenstates of the lower band, Eq.~\eqref{ghhg} yields
\begin{equation*}
e^{-2iNdk}\frac{\sin^2\frac{k_0 +k}{2}d}{\sin^2\frac{k_0 -k}{2}d}=
\frac{1+\sin(k_0 d_1+\frac{kd}{2})}{1-\sin(k_0 d_1-\frac{kd}{2})}.
\end{equation*}
Therefore the solutions obey
\begin{equation*}
\epsilon_\xi=\xi+i\frac{1}{2\pi}\ln\frac{1+\sin k_0 d_1}{1-\sin k_0 d_1},
\end{equation*}
and the eigenvalues are given by
\begin{subequations}
\begin{equation}
\omega_{-}(\xi)=\frac{\gamma_0 /2}{1-\cos k_0 d}(\sin k_0 d- \xi\frac{\pi}{N}\sin k_0 d_1),
\end{equation}
\begin{equation}
\gamma_{-}=\frac{\gamma_0}{4N}\cot(k_0 d_1)
\ln(\frac{1+\sin k_0 d_1}{1-\sin k_0 d_1}),
\end{equation}
\end{subequations}
which is the result given in the main text.

\bibliography{flatness.bib}